\documentclass[12pt, a4paper]{article}
\usepackage{amsmath}
\usepackage{amsthm}
\usepackage{amsfonts}
\usepackage{amssymb}
\usepackage{graphicx, rotating}
\usepackage{epstopdf}
\usepackage{epsfig}
\usepackage{latexsym}
\usepackage{graphicx}
\usepackage{color}
\usepackage{cite}
\usepackage{slashed}
\usepackage{float}
\usepackage[export]{adjustbox}

\usepackage{standalone}
\usepackage{physics}
\usepackage{textgreek}
\usepackage[dvipsnames]{xcolor}
\definecolor{rossos}{cmyk}{0,1,1,0.55}
\definecolor{bluscuro}{rgb}{0.15, 0.2, .85}
\definecolor{bluchiaro}{cmyk}{1,.3,0.,0.1}
\usepackage[colorlinks=True, citecolor=bluscuro, linkcolor=black, urlcolor=bluscuro,linktocpage]{hyperref}

\newcommand{\eq}[1]{Eq.~(\ref{#1})}

\newcommand{\nn}{\nonumber}

\newcommand{\be}{\begin{equation}}
\newcommand{\ee}{\end{equation}}
\newcommand{\bea}{\begin{eqnarray}}
\newcommand{\eea}{\end{eqnarray}}
\newcommand{\bc}{\begin{center}}
\newcommand{\ec}{\end{center}}

\newcommand{\bs}{M^2}

\newcommand{\M}{{\cal M}}
\newcommand{\Mh}{{\cal M}}

\newcommand{\doa}{\hspace{-2.5pt}\begin{array}{c}\smallfrown\\[-10pt]\smallsmile \end{array}\hspace{-2.5pt}}

\def\ee{\text{e}}
\def\ii{\text{i}}
\def\dd{\text{d}}

\usepackage[utf8]{inputenc}
\usepackage[english]{babel}
\usepackage{fancyhdr}
\usepackage{psfrag}
\usepackage{graphicx,wrapfig}
\usepackage[bf,footnotesize]{caption2}

\usepackage[sort&compress,numbers, merge]{natbib}

\usepackage{amsthm,bm}
\usepackage{gensymb}

\usepackage{physics}
\usepackage{array}
\usepackage{tcolorbox, mathtools}
\usepackage{soul}
\usepackage{feynmp-auto}

\usepackage[section]{placeins}

\begin{document}

\vspace*{-2cm}
\begin{flushright}
\vspace*{2mm}
\today
\end{flushright}

\begin{center}
\vspace*{15mm}

\vspace{1cm}
\large \bf 
The EFT Bootstrap at Finite $M_{PL}$ 
\vspace{1.4cm}

{ Carl Beadle$\,^{a}$, Giulia Isabella$\,^{a,b}$, Davide Perrone$\,^{a}$, Sara Ricossa$\,^{a}$, Francesco~Riva$\,^{a}$, Francesco Serra$\,^{a,c,d}$}

 \vspace*{.5cm} {\it
$^a$ D\'epartment de Physique Th\'eorique, Universit\'e de Gen\`eve\\
$^b$ Mani L. Bhaumik Institute for Theoretical Physics, Department of Physics and Astronomy, University of California Los Angeles\\
$^c$ Scuola Normale Superiore, Pisa\\
$^d$ Department of Physics and Astronomy, Johns Hopkins University
\vspace*{.2cm} 
}
\end{center}

\vspace*{10mm} 
\begin{abstract}\noindent\normalsize

\noindent
We explore the impact of loop effects on positivity in effective field theories emerging in the infrared from unitary and causal microscopic dynamics. Focusing on massless particles coupled to gravity, we address the treatment of forward-limit divergences from loop discontinuities and establish necessary conditions for maintaining computational control in perturbation theory. 
While loop effects remain small, ensuring consistency in our approach leads to a significant impact on bounds, even at tree level.
%While loop effects remain small, ensuring consistency in our approach leads to significant deviations from standard bounds, even at tree level.

\end{abstract}

\vspace*{3mm}
\newpage
\tableofcontents

\newpage

\section{Motivation}

Effective Field Theories (EFTs) are essential tools for describing physics at the boundary between measurable phenomena and the unknown. Universality—the principle that different high-energy theories converge to the same EFT at low energies—allows EFTs to be applied agnostically to a wide range of systems, including gravity, physics beyond the Standard Model and pion physics. However, this universality also limits EFTs' intrinsic predictive power. Strictly speaking, the only robust predictions within an EFT are its low-energy analytic structures, governed by the interplay of calculable loop effects and unknown Wilson coefficients.

Remarkably, even minimal ultraviolet (UV) assumptions such as unitarity and causality can endow EFTs with predictive and testable features. Dispersion relations exploit the analytic properties of amplitudes to establish IR-UV \emph{positivity} bounds, which identify EFTs compatible with unitary UV completions~\cite{Adams:2006sv,Arkani-Hamed:2020blm,deRham:2017avq,Bellazzini:2020cot,Sinha:2020win,Tolley:2020gtv,Caron-Huot:2020cmc,Caron-Huot:2021rmr,Donoghue:1995cz}.
These bounds are particularly significant in quantum gravity, where UV completions are sparse and mostly inaccessible to direct testing~\cite{Adams:2006sv, Caron-Huot:2021rmr,Bellazzini:2015cra, Cheung:2016wjt,Bonifacio:2018vzv, Hamada:2018dde, Bellazzini:2019xts,  Melville:2019tdc, Tokuda:2020mlf,Bern:2021ppb, Caron-Huot:2021enk,Arkani-Hamed:2021ajd, Bellazzini:2021shn,Caron-Huot:2022ugt,Caron-Huot:2022jli,Chiang:2022jep,Bellazzini:2022wzv,Henriksson:2022oeu,Herrero-Valea:2020wxz,Herrero-Valea:2022lfd, Edelstein:2021jyu,Serra:2022pzl, Bellazzini:2023nqj, Chiang:2022jep,Haring:2022cyf}. 
A significant breakthrough in this area is the development of a systematic approach to addressing the Coulomb singularity in graviton exchange. This method involves smearing dispersion relations with momentum-dependent kernels, as introduced in~\cite{Caron-Huot:2021rmr} and further refined in ~\cite{Beadle:2024hqg} by fully departing from the forward limit, allowing for a more robust and effective treatment of these singularities.

Positivity bounds have far-reaching implications, including an EFT-based rationale for the weak gravity conjecture~\cite{Bellazzini:2019xts,Alberte:2020jsk}, constraints on ultra-soft interactions~\cite{Bellazzini:2016xrt} and the exclusion of massive gravitons~\cite{Cheung:2016yqr,Bellazzini:2017fep,deRham:2018qqo,Bellazzini:2023nqj}. 
They delineate an \emph{EFT swampland}, defining how quantum field theories behave at long distances and serving as a foundation for understanding the swampland in string theory.

Most efforts to map the allowed parameter space of EFTs have assumed a \emph{tree-level} framework, where the amplitude's non-analytic features arise solely from high-energy exchanges above the EFT cutoff. This approach neglects calculable non-analyticities within the EFT. In this article, we systematically examine the infrared structure of EFT amplitudes at one-loop order, focusing on $2\to 2$ scattering of massless scalar particles interacting via gravity in various spacetime dimensions. Using on-shell methods, we compute the amplitudes and investigate how loop effects influence positivity bounds. While loop corrections to positivity are not new, see for instance~\cite{Arkani-Hamed:2020blm,Bellazzini:2020cot, Bellazzini:2021oaj, Acanfora:2023axz, Ye:2024rzr}, their interplay with gravity introduces novel questions and complexities, including Coulomb singularities, which requires a careful treatment of the S-matrix \cite{Strominger:2017zoo,Giddings:2011xs,Ware:2013zja} and the use of smeared dispersion relations.
In this work, we address these in different frameworks, including fixed-$t$ (FT) dispersion relations~\cite{Caron-Huot:2021rmr,Beadle:2024hqg} as well as crossing symmetric dispersion relations (CS)~\cite{Sinha:2020win,Chowdhury:2022obp,Li:2023qzs,Song:2023quv,Berman:2024kdh,Zahed:2022wuy,Chowdhury:2021ynh}.

We address many important subtle points in this study.
First of all, at sufficiently low energies, loop effects due to the most relevant operators dominate over less relevant ones. This dominance impacts positivity bounds, as calculable IR contributions can overshadow constraints on higher-order coefficients, weakening their utility for shaping the EFT swampland. %~\cite{Bellazzini:2020cot}.
Such effects also yield surprising results, like negative corrections to the coefficients of irrelevant operators~\cite{Bellazzini:2020cot} and to the charge-to-mass ratio of extremal black holes in theories with photons and gravitons, which are pertinent to the weak gravity conjecture~\cite{Cheung_2014, Arkani-Hamed:2021ajd}.
We find that in scalar-gravity systems, the dominant gravitational IR effects scale as $\kappa^4 s^3 \sqrt{-t}$ 
or $\kappa^4 s^3 \log(-t)$ in $d=5,6$, where the calculation is particularly well defined. 
For sufficiently small $s$ and $t$, these contributions can surpass loop-factor suppression and dominate over less relevant contact interactions. Most importantly, the presence of massless particles introduces singularities that complicate the use of dispersion relations, especially in the forward limit. This complexity renders the concept of tree-level and weak coupling effectively meaningless in such scenarios.

Secondly, loop corrections also modify the analytic structure of amplitudes. While tree-level amplitudes are real and dispersion relations can isolate specific energy-expansion coefficients, loop corrections introduce branch cuts. These discontinuities, governed by unitarity, depend on the ensemble of all couplings, necessitating assumptions about EFT convergence before deriving bounds.

Finally, it remains uncertain whether gravity will be UV completed at weak or strong coupling. In the case of strong coupling, the effects explored in this work offer a qualitative insight into how the EFT swampland deviates from the idealized tree-level framework.

This study represents a foundational step toward bridging the gap between the tree-level results of the positivity program and the fully non-perturbative S-matrix bootstrap, which relies on ansatz-driven methods~\cite{Guerrieri:2021ivu}. The interplay between loop- and tree-level effects becomes particularly significant in the presence of multiple particle species, playing a crucial role in unraveling the path to a consistent UV completion of the theory~\cite{Caron-Huot:2024lbf}.

This article is structured as follows. In section~\ref{sec:EFTAmplitude} we compute 1-loop effects in the amplitude of spin-0 particles interacting through gravity.
We discuss how these effects enter the different dispersion relations in section~\ref{sec:IRE}.
In section~\ref{sec:BoundsMod} we show how the bounds are modified beyond tree-level.
In the Appendices we provide more  detailed expressions 
on the numerics~\ref{app:smearing}.\\

\textbf{Note added}: After the completion of this work, we learned of an upcoming work with overlapping results~\cite{Graviton_2025}. We thank the authors for sharing the draft and for important discussions.

%%%%%%%%%%%%%%%%%%%%%%%%%%%%%%%%%%%%%%%%%%%%%%%
\section{The  Structure of EFT Amplitudes}\label{sec:EFTAmplitude}

\subsection{Tree-Level}
\label{sec:tree}

 We focus on the $2\to 2$ scattering amplitude for  exact (massless) Goldstone bosons in generic $d$-dimensions in a theory with a mass-gap $M$, such that at sufficiently small energy $E\ll M$ the theory is weakly coupled and well described by an effective  Lagrangian with interactions organised in a derivative expansion.
 We insist on the Goldstone nature of the spin-0 particles to naturally justify their masslessness that we assume throughout.  
 At tree-level, the  $2\to 2$ amplitude can be written as  $\M_{\text{tree}}=\mathcal{M}_{\text{tree}}^{\text{EFT}} + \mathcal{M}_{\text{tree}}^{\text{grav}}$ with one part associated with contact interactions,
\begin{equation}\label{amptreeXY}
	\mathcal{M}_{\text{tree}}^{\text{EFT}} = 
\sum_{n\geq 2,q\geq 0} g_{n,q} \left(\frac{s^2 + t^2 + u^2}{2}\right)^{\frac{n-3q}{2}}\cdot \left(s  t u\right)^{q}
\end{equation}
and one mediated by gravity,
\begin{equation}\label{eqap:GravTree}
	\mathcal{M}_{\text{tree}}^{\text{grav}} =  \kappa^2 \left( \frac{t u}{s} + \frac{u s}{t} + \frac{s t}{u} \right), \;\;\; \text{where} \;\; \kappa^2 \equiv \frac{1}{M_{Pl}^{d-2}}
\end{equation}
denotes the gravitational constant in $d$-dimensions,  and $M_{Pl}$ the Planck scale.\footnote{A constant term in the amplitude \eq{amptreeXY} and a pole associated with  the scalar exchange  are forbidden by the Goldstone-Boson shift-symmetry.} 

It is illustrative to keep track of the size of the various terms in situations where the EFT is dominated by a single scale $M$ and a dimensionless coupling $g$, such that, 
\begin{equation}\label{eq:PowerCountingHeuristic}
    g_{n,m}\sim \frac{g^2}{M^{2n+d-4}}\,.
\end{equation}
This relation holds in simply weakly coupled UV models, but also captures  important features at strong coupling~\cite{Giudice:2007fh}. 
This expression helps to separately keep track of the EFT energy expansion controlled by $E/M\ll1$ and the EFT loop expansion, controlled by $g^2/(4\pi)^{d/2}\ll 1$. For gravity instead it's roughly controlled by $\kappa^4 M^{2d-4}/(4\pi)^{d/2}$. 

Gravity is always more relevant than all other EFT interactions and dominates at sufficiently small  energies. Moreover, even loops of gravity can dominate over certain EFT interactions.

This happens for  irrelevant EFT operators with  $n\geq d/2$,  at energies
\begin{equation}\label{eqNDA}
    \frac{E}{M} \lesssim \frac{E_*}{M}\equiv \left[( (4 \pi)^{d/2} \, g )^{-2} \left(\frac{M}{M_{Pl}}\right)^{2(d-2)} \right]^{\frac{1}{2n-d}} \, .
\end{equation}
 Therefore the tree-level EFT is a valid approximation only within a window at sufficiently high energy where gravity loops are negligible, but with energies sufficiently small that the EFT description holds, ${E_*}/{M}\ll {E}/{M} \ll 1 $. For larger $n$, this window shrinks and disappears, meaning that for statements on $n\gg d/2$ couplings, loops are always important.
 Furthermore dispersion relations, by construction, are sensitive to the amplitude at all energies, in particular also in the region of \eq{eqNDA}, where loops dominate. In the rest of this section we compute these effects more precisely and study how they appear in dispersion relations.

\subsection{IR Effects}
\label{sec:loopsIR}
We are interested in how higher order effects modify the structure of \eq{amptreeXY} by altering the analytic structure in both $s$ and $t$, as well as how this impacts dispersion relations.  
The inclusion of long-range interactions has different consequences in the study of scattering amplitudes \cite{,Giddings:2011xs,Strominger:2017zoo,Ware:2013zja}.
First of all, it is well known that IR divergences in diagrams with a fixed number of external legs cancel against divergences  in the real IR radiation, when appearing in the total (inclusive) cross section. 
The contribution to the total cross-section from the real emission of gravitons diverges in $d\leq 4$ due to collinear/soft effects, which also implies that the amplitude is not well-defined for any exclusive process (e.g. $2\to2$ scattering).
To make this more concrete, in this work we study the case of $d>$4, where the problem is absent and the phase-space integral of collinear radiation is finite. We discuss the case $d=4$ separately in sec.~\ref{sec:d4}.
Technically, the study of $d>4$ in the context of gravity is made possible by the convergence of the partial wave expansion, which diagonalises the unitary property of the S-matrix.
Its coefficients are given by the integrals,
\begin{equation}\label{eq:PWs}
f_\ell(s)=\mathcal{N}_d\int_{-s}^{0} \frac{dt}{s}\left({4\, tu}/{s^2}\right)^{(d-4)/2}{\cal P}_\ell\left(1+2{t}/{s}\right)\M (s,t)\,,\qquad \text{with}\quad \mathcal{N}_d =\frac{ (16\pi)^{\frac{2-d}{2}}}{ \Gamma\left(\frac{d-2}{2}\right)}\,,
\end{equation}
where 
${\cal P}_\ell(x)=\,_2F_1\left(-\ell, \ell + d - 3, (d - 2)/2, (1 -x)/2\right)$
are  Gegenbauer polynomials, and $u=-s-t$.
This allows us to write the amplitude in the partial wave expansion as,
\begin{equation}
 \label{Eq:partial_wave_expansion}
  \Mh(s,t)= \sum_{\ell=0}^\infty n_\ell^{(d)}\, {\cal P}_\ell\left(1+\frac{2t}{s}\right) f_\ell(s)\,,
\end{equation}
where 
\begin{equation}
   n_\ell^{(d)} = \frac{(4\pi)^{\frac{d}{2}}(d+2\ell-3)\Gamma(d+\ell-3)}{\pi \, \Gamma\left(\frac{d-2}{2} \right)\Gamma(\ell+1)} \,,
\end{equation} 
and $\ell$ runs over even integers for identical scalars.

For gravity in $d=4$ this expansion does not converge due to the pole of the amplitude ~$\sim 1/t$.
 In $d\geq$ 5 it converges but only as a distribution~\cite{Kravchuk:2020scc,Haring:2022cyf} appearing within integrals over given measures.
In other words,
the amplitude and the dispersion relations that follow will have to be {smeared} with weight-functions in $t$, rather than being thought of as functions of $t$.

After the above IR ambiguities have been addressed, massless particles still leave their imprint via finite computable loop effects.
These are physical predictions of the theory, they also exist in $d>4$ and  would plausibly survive in an IR safe definition of the S-matrix. 
We are interested in  the ones that modify the analytic structure of the amplitude as $s,t\to 0$, since  these have an impact on how dispersion relations can be used.
We refer to these generically as \emph{IR effects}, and we seek to identify the most relevant ones.

%%%%%%%%%%%%%%%%%%%%%
\subsection{Finite IR Effects in the theory of a scalar and gravity}\label{sec:loops}

For our purpose, the most important IR effects arise at 1-loop, because they represent the first qualitative modification of the amplitude analytic structure w.r.t. tree-level.
As discussed above 1-loop effects can  in principle be relatively large under certain circumstances, without necessarily implying a breakdown of the perturbative expansion. 
Indeed, higher loops introduce more powers of energy and soften the $s,t\to 0$ behaviour, thus playing less of an important role in the context of positivity bounds.
In this section we compute these 1-loop effects. Although parts of these already appear in the literature, we present a systematic study in general $d$ and including EFT couplings.  We work in dimensional regularisation $d=D-2\epsilon$ for integer $D$, which is recovered by assuming that all couplings are defined in integer dimensions.

%%%%%%%%%%%%%%%%%%%%%

The one-loop contribution to the amplitude can be divided into three pieces,
\begin{equation}\label{amp-full}
    \mathcal{M}_{1\text{-loop}}=\mathcal{M}_{1\text{-loop}}^{\text{EFT-EFT}}+\mathcal{M}_{1\text{-loop}}^{\text{grav-EFT}}+\mathcal{M}_{1\text{-loop}}^{\text{grav-grav}},
\end{equation}
where the subscripts denote the power counting both in terms of gravitational and EFT couplings.
Each piece can be projected onto a basis of scalar one-loop integrals~\cite{Passarino:1978jh,Ellis_2012}.
As we are looking at the $2\to2$ amplitude, they are limited to bubble, triangle and box integrals, but because all states are massless the contribution from all triangle integrals can themselves be projected onto bubble integrals using integration by parts (IBP) identities \cite{Tkachov:1981wb,Chetyrkin:1981qh, Smirnov:2012gma}.
This leaves,
\begin{equation}\label{eqap:PVreduction}
	\mathcal{M}_{1\text{-loop}} = \sum_{i=\bigcirc,\Box} c_i \, \mathcal{I}_i  \, .\\
\end{equation}
where the $c_i$ are rational functions of the kinematic variables and,
\begin{eqnarray}
    \label{eq:MIC}
        \mathcal{I}_{\bigcirc}(t) &=& \mu^{2\epsilon}\int \frac{d^d l}{\left( 2 \pi \right)^d} \frac{1}{l^2} \, \frac{1}{\left( l + p_1 + p_3 \right)^2}=\frac{i\mu^{2\epsilon}\,\Gamma \left(2-\frac{d}{2}\right) \Gamma \left(\frac{d}{2}-1\right)^2}{(4\pi)^2\,\Gamma (d-2)} \,\left(-\frac{t}{4\pi}\right)^{d/2-2},\\
        \mathcal{I}_{\Box}(s,t) &=& \mu^{2\epsilon}\int \frac{d^d l }{\left(2 \pi\right)^d} \frac{1}{l^2} \, \frac{1}{ (l+p_1)^2} \, \frac{1}{(l+p_1+p_3)^2} \, \frac{1}{(l-p_2)^2}=
        -\frac{i\,\mu^{2\epsilon}\,\Gamma{\left(2-\frac{d}{2}\right)}\Gamma{\left(\frac{d}{2}-2\right)}^2}{(4\pi)^2\,s t\,  \Gamma{(d-4)}}
      \nn
      \\ &&\label{eq:MIS}
       \times \Big[ \left(-\frac{s}{4\pi\mu^2}\right)^{\frac{d}{2}-2}\,_2F_1\left(1,\frac{d}{2}-2,\frac{d}{2}-1,1+\frac{s}{t}\right)
    +(s\leftrightarrow t)\Big] \,.
\end{eqnarray}
These expression are most often found in the literature having applied dimensional regularisation in the $d=4$ case \cite{Ellis_2008,Weinzierl2022}, or in generic dimension $d$ \cite{Smirnov2005}. 
The hypergeometric function in \eq{eq:MIS} will play an important role for us because of its analytic structure; explicitly,
\begin{equation}\label{eq:expsing}
    _2F_1\left(1,\frac{d}{2}-2,\frac{d}{2}-1,1+\frac{s}{t}\right)=\left\{\begin{array}{ll} 1& d=4\\
    \frac{\tanh ^{-1}\left(\sqrt{1+s/t}\right)}{\sqrt{1+s/t}}& d=5\\
    -\frac{\log (-s/t)}{1+s/t}&d=6\\
    \end{array}\right.\,.
\end{equation}

To determine the factors $c_i$  in \eq{eqap:PVreduction} we  use reverse unitarity \cite{Anastasiou:2002yz,Anastasiou:2002qz,Anastasiou:2003yy,Anastasiou:2015yha}.
Unitarity dictates that the one-loop integrand discontinuity is given by the product of two tree-level amplitudes and by projecting said discontinuity onto that of the above $1$-loop scalar integrals in \eq{eq:MIC} and \eq{eq:MIS}, we can then extract $\mathcal{M}_{1\text{-loop}}$.
In any specific number of dimensions, this type of procedure determines the 1-loop amplitude only up to possible rational terms that do not have the above-mentioned singularities. 
However, since we consider general $d$-dimensional integrands, these rational terms are also uniquely determined; they contribute to non-analitycities in other dimensions and are therefore picked up by our procedure~\cite{Britto:2010xq}.
IBP identities are applied with the help of the `LiteRed' package \cite{lee2012presenting,Lee:2013mka}.

\begin{figure}[H]
\begin{center}
\includegraphics{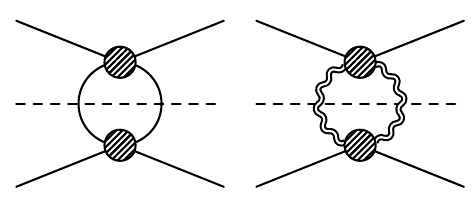}

\end{center}
\caption{\emph{\footnotesize
Non-trivial cuts used for the matching in \eq{eqap:PVreduction}.
LEFT: cuts for $1$-loop diagrams with scalar propagators in the `$t$-channel'. RIGHT: cuts for $1$-loop diagrams with gravitons.
The grey blobs denote all  possible tree-level interactions associated with either graviton exchange or insertion of an EFT $4$-point interaction.
}}
\label{fig:oneloopdiagrams}
\end{figure}

The one-loop amplitude is then reduced  to a weighted sum of box and bubble diagrams of the form,
\begin{equation}
\begin{split}
	 \mathcal{M}_{1-\text{loop}} & = f_1\left( s, t \right) \mathcal{I}_{\Box} \left( s, t \right) + f_2\left( u, s \right) \mathcal{I}_{\Box} \left( u, s \right) + f_3\left( t, u \right) \mathcal{I}_{\Box} \left( t, u \right)  \\
	&+   g_1\left(u, s\right) \mathcal{I}_{\bigcirc} \left( t \right)+ g_2\left(s,t\right) \mathcal{I}_{\bigcirc} \left( u \right) +g_3\left(t,u\right) \mathcal{I}_{\bigcirc} \left( s \right)  , \label{eqf1f2f3}\\
\end{split}
\end{equation}
where $f_i$ and $g_i$ are functions of $d$ and of the external momenta. They are in principle independent of eachother.
However, since the particles scattered are identical scalars, crossing symmetry implies that the functions in \eq{eqf1f2f3} are  related by,
\begin{align}
    f_1\left( x,y \right)&=f_2\left( x,y \right)=f_3\left( x,y \right)\equiv f_d(x,y)\, ,\nn \\
    g_1\left( x,y \right)&=g_2\left( x,y \right)=g_3\left( x,y \right)\equiv g_d(x,y)\, ,\label{eq:fdgd}
\end{align}
and are reduced to two independent functions $f_d,g_d$.
This also implies that it is sufficient to match the discontinuities of an individual cut, which we choose to be the $t$-channel cut, as illustrated in  Fig.~\ref{fig:oneloopdiagrams} -- the discontinuities in the other channels will then be reproduced by the identities~\eq{eq:fdgd}. As shown in the figure,
there are two contributions to this cut, one obtained by cutting internal scalar legs, and a second by cutting internal graviton legs.

\subsubsection{EFT interactions}

First we focus on the interactions without gravity, $\mathcal{M}_{1\text{-loop}}^{\text{EFT-EFT}}$.
The tree-level EFT has only 4-point contact interactions \eq{amptreeXY}. Because of this it is clear that at $1$-loop level, there can only be a projection onto scalar bubble integrals, implying $f_d=0$.
At small enough energies, the leading effects within the EFT should come from the couplings $g_{n,m}$'s in \eq{amptreeXY} that are labelled by the smallest integers, and dominated by the term $\propto g_{2,0}^2$, followed by less relevant terms from the mixing of $g_{2,0}$ and other Wilson coefficients.

Performing the $t$-channel cut on the scalar legs and counting the powers of Wilson coefficients allows us to isolate some of the contributions; the others will be reproduced via crossing symmetry.
Loops involving arbitrary EFT coefficients share a common, dimension dependent factor, which stems from the scalar bubble integral~$\mathcal{I}_{\bigcirc} (t )$.  

The most relevant contributions can be written as,
\begin{eqnarray}\label{eq:1left2}
 \mathcal{M}_{1\text{-loop}}^{\text{EFT-EFT}} &=&  \frac{\, t^2}{2^{4} \left(d^2-1 \right)} \mathcal{I}_{\bigcirc} (t ) \Bigg[  g_{2,0}^2 \, \left(4 s u- \frac{3}{2} d \, (3 d+2) t^2\right)  \\
 && +  g_{2,0} \, g_{3,1}  \, t \left(((2-3 d) d+8) t^2-8 s u\right)  + g_{3,1}^2 \, \frac{t^2}{2}  \left(8 s u-(d-2) d t^2\right)  \nn\\
 &&   - g_{2,0} \, g_{4,0} \,\frac{3 t^2 \left(\left(9 d \, (d+2)^2+32\right) t^2-16 (d+4) s u\right)}{4 (d+3)} \,   \nn +\cdots \Bigg] +\left(t \leftrightarrow s,u \right) .\nn
\end{eqnarray}
This matches Ref.~\cite{Bellazzini:2021oaj} in the $d \rightarrow 4$ limit.
Polynomial pieces in \eq{eq:1left2} are resorbed into the definitions of the renormalised EFT coefficients.
Higher order terms, denoted by dots in \eq{eq:1left2}, become less and less relevant but are systematically calculable.

For $ \mathcal{M}_{1\text{-loop}}^{\text{EFT-EFT}}$, the function $\mathcal{I}_{\bigcirc}(t)$, and its crossed counterparts $\mathcal{I}_{\bigcirc}(s),\mathcal{I}_{\bigcirc}(u)$, are responsible for the amplitude's non-analyticities.
These are associated entirely to the factor $(-t)^{d/2}$ in \eq{eq:MIC}, multiplied by a polynomial in the crossing symmetric combinations $su,t^2$, and similarly for the other channels. In even dimensions these will lead to  logarithmic discontinuities, while in odd dimensions to  square-root ones.

\subsubsection{Mixed EFT-Gravity Interactions}
Loops involving both gravity and EFT interactions, also have the property that they project onto scalar bubble integrals only; because of this, they have a structure similar to EFT-EFT effects.
The most relevant pieces are,
\begin{eqnarray}
\mathcal{M}_{\text{1-loop}}^{\text{grav-EFT}} &=& 
    \kappa^2\frac{t^2}{2^{4} \left(d^2-1 \right)} \mathcal{I}_{\bigcirc} (t ) \Bigg[ g_{2,0} \, \frac{6 \left((d-2) \left(9 (d-2) d^2+32\right) t^2-8 (d (5 d+2)-8) s u\right)}{(d-4) (d-2) t} \, \nn
\\ &&
    + g_{3,1} \, 6 \left(\frac{8 (d (5 d+2)-8) s u}{(d-4) (d-2)}+d \, (3 d+2) t^2\right)  
\\ &&
     +g_{4,0} \Bigg[ \, \frac{3 \left((d-2) d (3 d+4) \left(9 d^3-36 d+128\right) t^4+2048 (d+1) (d+3) s^2 u^2\right)}{2 (d-4) (d-2) d (d+3) t} \nn \\
  && \quad \quad \quad +\frac{3 \left(-16 d (d (3 d (5 d + 22) + 64) - 32) s t^2
    u \right)}{2 (d-4) (d-2) d (d+3) t}
     \Bigg] +\cdots \Bigg] +\left(t \leftrightarrow s,u \right) \nn 
\label{eq:1lEFTg}
\end{eqnarray}

 The analytic structure of these effects is  similar to that of EFT-EFT diagrams discussed above.
 Notice that in  $d>2$ dimensions, the apparent $ 1/t $ pole in the first line is cancelled by  positive powers of $t$ in $\mathcal{I}_{\bigcirc} (t )$. 
 At low enough energy, these effects dominate over the ones in the previous paragraph, since gravity is a more relevant interaction.

\subsubsection{Gravity Only}
\begin{figure}[H]
\begin{center}
\includegraphics{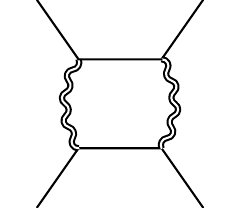}
\end{center}
\caption{\it \footnotesize
One of many diagrams contributing  to the $1$-loop amplitude at order $\kappa^4$. 
}
\label{fig:egloopdiagram}
\end{figure}

The largest loop effects in the IR are  associated with diagrams involving only gravitational interactions. An example of such diagrams is shown in Fig.~\ref{fig:egloopdiagram}, which contributes to the cut represented in Fig.~\ref{fig:oneloopdiagrams}.
To compute the effects from  diagrams of this class using reverse unitarity, we must cut the internal graviton legs in the $t$-channel and sum over the different graviton polarisations.
For this we need the tree-level scattering amplitude of spin-0 particles and gravitons in generic dimension $d$ \cite{Bjerrum_Bohr_2019}.
This can be compactly written as,
\begin{equation}\label{eqap:treelevelgrav}
	\mathcal{M}_{\phi h \to \phi h} = \frac{\kappa^2}{s t u} \left[ 2 s \left(\epsilon_1\cdot p_4\right)\left(\epsilon_3\cdot p_2\right)  + 2 u \left(\epsilon_1\cdot p_2\right)\left(\epsilon_3\cdot p_4\right) + s u \left( \epsilon_1 \cdot \epsilon_3 \right)\right]^2,
\end{equation}
which in $d=4$ can be recast in spinor-helicity language, simplifying to
\begin{equation}\label{eqap:spheltreegrav}
	\mathcal{M}_{\phi h \to \phi h} \Bigg|_{d=4} = \frac{\kappa^2}{s t u} \left[ 1 | 2 | 3 \right. \rangle ^4 .\\
\end{equation}.

The sum over graviton helicities  is given by the full momentum-dependent graviton  propagator, suitable to be employed with this tree-level amplitude~\cite{Kosmopoulos:2020pcd}:
\begin{equation}
    \sum_{\lambda}\epsilon_\lambda^{\mu\nu}(\ell)\epsilon_\lambda^{\rho\sigma}(-\ell)
    = \frac{1}{2}\left(P^{\mu\rho}P^{\nu\sigma}+P^{\mu\sigma}P^{\nu\rho}
    - \frac{2}{d-2}P^{\mu\nu}P^{\rho\sigma}
    \right),
\end{equation}
where,
\begin{equation}
    P^{\mu\nu} = \eta^{\mu\nu} - \frac{\ell^\mu q^\nu + \ell^\nu q^\mu}{\ell\cdot q},
\end{equation}
and $q$ is a reference momentum.

Repeating the matching procedure for the one-loop gravity corrections allows us to write $\mathcal{M}_{1-\text{loop}}^{\text{grav-grav}}$ in the form of \eq{eqf1f2f3} with\footnote{We thank J. Parra-Martinez and C.-H. Chang for pointing out an error in a previous version of this formula.},
\begin{equation}\label{eq:symmetricmandelsfuncs}
	f_d\left(x, y \right) \equiv \kappa^4(x^4 + y^4)\,, \quad\quad 
	g_d\left( x, y \right) \equiv \kappa^4 \left( x^2 +y^2 \right) r_1 \left( d \right) +\kappa^4 \,x\, y\,  r_2 \left(d\right)\,.
\end{equation}
The dimension specific rational functions are themselves given by,
\begin{equation}\label{eq:rationalfuncs}
\begin{split}
	r_1 \left( d \right) &\equiv \frac{d^6-11 d^5-562 d^4+2820 d^3-2792 d^2-2848 d+3584}{64 \left(d-4\right)\left(d-2 \right)  \left(d^2-1\right)}, \\
	r_2\left(d\right) &\equiv \frac{d^6-27 d^5+74 d^4-232 d^3+648 d^2+496 d-576}{32 (d-4) (d-2) (d^2-1)}. \\
\end{split}
\end{equation}
Contrary to loops involving EFT interactions, here we inherit the non-analyticities of the box integral~$\mathcal{I}_{\Box}$. 
The hypergeometric function $_2F_1 \left( 1, \frac{d}{2}-2 , \frac{d}{2}-1 ; 1 + z \right)$, with $z=s/t$ in the  box integral \eq{eq:MIS} has a branch cut which extends from $z = 0$ to real infinity, see \eq{eq:expsing}, but is otherwise analytic everywhere.
Therefore, at fixed-$t$, the amplitude contains a branch cut on the real axis.
As we will review in section~\ref{sec:IRE}, the coefficient of such a discontinuity enters dispersion relations with an \emph{arbitrary number of subtractions}, when the contour of integration is taken across the $\mathcal{I}\text{m} \left[s\right] = 0$ axis.
In turn, this coefficient has $t\to 0$ singularities.
We find that the most singular such pieces are, \footnote{In the Regge limit $\frac{t}{s}\ll 1$,  this leading contribution  arises from the box diagram, and is consistent with the first iteration of the tree-level graviton exchange relevant in the eikonal approximation, see e.g. \cite{Amati:1987uf,Amati:1987wq,Amati:1990xe,Amati:1992zb}.}
\begin{equation}\label{eq:singularities}
	\text{Disc} \left[\ii \mathcal{M}_{1-\text{loop}}^{\text{grav-grav}} \right] \underset{t \ll s}{\sim} \kappa^4 \left( - t \right)^{\frac{d - 6}{2}} \, s^3 \begin{cases} \log \left( -t \right) & d \text{ is even,} \\ 1 & d \text{ is odd} \end{cases} 
\end{equation} 
where we have regulated the integrals using the dimensional regularisation in the $\overline{\text{MS}}$  scheme, which we employ throughout.
This is an important result that reveals how beyond-tree-level dispersion relations can be employed.
In particular it shows that in dimensions $d\leq 6$ all dispersion relations  diverge in the forward limit, while in $d\leq 8$ the dispersion relations' first derivative in $t$ will diverge.

Lastly, from the non-analyticities of the one-loop amplitudes reported above, we can easily extract the running of the Wilson coefficients. In particular, we observe that $g_{4,0}$ runs for $d\leq 8$,  $g_{3,1}$ for $d\leq 6$ and $g_{2,0}$ for $d\leq 4$, as expected by naive dimensional analysis arguments~\cite{Anber:2011ut}.

%%%%%%%%%%%%%%%%%%%%%%%
\subsection{In 4 dimensions}\label{sec:d4}
As mentioned above, S-matrix elements in $d=4$ are affected by IR divergences when massless particles are considered. 
After regularisation, these divergences can be cancelled by considering inclusive observables or different notions of asymptotic states, see e.g.~\cite{Kinoshita:1962ur,Lee:1964is,Hannesdottir:2019opa,Hannesdottir:2019umk}, or resummed when the IR cutoff has a physical meaning, in the case of IR sensitive observables, see e.g. \cite{Bauer:2000ew,Bauer:2003pi,Bosch:2004th}.

In this section, we compute the $d=4$ IR divergences that appear for generic kinematic configurations and we compare them with the $t\to 0$ singularities of Eq.~\eqref{eq:singularities}.
This comparison makes it clear that the $t\to 0$ singularities affecting the dispersion relations are not captured by kinematic-independent IR divergences that might be resorbed by redefinition of the asymptotic states. 
Rather, the contributions in Eq.~\eqref{eq:singularities} are the result of the dynamical properties of gravitational scattering in $t=0$.

To see this we can compute explicitly the kinematic-independent one-loop IR divergences of a scattering amplitude in the presence of gravity.  
In dimensional regularisation, $d=4-2\epsilon$, the IR divergences take the form of poles around $\epsilon=0$ whose residues are non-polynomial functions of the Mandelstam invariants.
At one-loop, these divergences will be of two kinds. 
The first kind of kinematic-independent divergences are those corresponding to massless bubble loops on external legs, contributing to the so-called collinear anomalous dimension. 
Since the on-shell computation of the one-loop amplitude uses connected tree-level amplitudes as building blocks (see for example Sec.~\ref{sec:loops}), the result of Eq.~\eqref{eq:singularities} is not affected by these bubble loops on the external legs.
The second kind of kinematic-independent IR divergences are those corresponding to the exchange of one soft particle between two external legs, contributing to the Sudakov double logs and to the so-called cusp anomalous dimension, see e.g. \cite{Sterman1978,Sterman1978b,Becher:2009cu,Bern:2020ikv}.
Indeed, aside from bubble loops on the on-shell legs, kinematic-independent IR divergences can only arise when three consecutive propagators are on-shell. 
In this case the loop will be of the form:
\begin{align}
    \frac{d^4p}{p^2(k_1+p)^2(k_2-p)^2}\sim\frac{d^4p}{p^2\,p\!\cdot\!k_1\,p\!\cdot\!k_2}\;,
\end{align}
which gives a divergent integral when $p^2=k_1^2=k_2^2=0$, regardless of the direction of $k_1$ and $k_2$. The only case in which the three legs are on-shell regardless of the kinematics will be when the loop arises from the exchange of a soft particle between two external legs.
Therefore, we only have to compare the $t\to0$ singularity of Eq.~\eqref{eq:singularities} to the contributions from exchanges of soft particles between external legs of a tree-level amplitude. 
\begin{figure}[H]
\begin{center}
\includegraphics{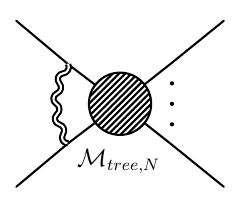}
\end{center}
\caption{\it Example of one-loop diagram contributing to the cusp anomalous dimension in an $N$-point scattering amplitude. 
The graviton internal line is taken to have soft momentum. External particles are taken to be on-shell.}
\label{fig:softgrav}
\end{figure}

In the case of scattering of shift-symmetric scalars coupled to gravity, at one-loop only a graviton can be exchanged between external legs -- as there are no three-point scalar self-interactions. 
The one-loop IR divergence will be proportional to the tree-level scalar $N-$point amplitude $\mathcal{M}_{\text{tree},N}$ and will have the following form in dimensional regularisation, see for example \cite{Dunbar:1995ed}:
\begin{align}
	\frac{i r_{\Gamma}}{(4\pi)^{2-\epsilon}}\frac{\kappa^2}{4\epsilon^2}\left (\sum_{i\neq j}^{N}(-s_{ij})^{1-\epsilon}\right ) \mathcal{M}_{\text{tree},N}\;,
\end{align}
where: $r_{\Gamma}=\Gamma^2(1-\epsilon)\Gamma(1+\epsilon)/\Gamma(2-\epsilon)$ and the sum is performed on the Mandelstam invariants for $N$ particle scattering in the unphysical region $s_{ij}<0\;,\; \forall \,i,j$.

Expanding at leading order for small $\epsilon$, for $N=4$ particles, we find:
\begin{align}
\left (\frac{1}{\epsilon}\frac{-s \log(-s)}{(4\pi)^2}+\frac{s\log (-s)
}{32\pi^2}\left (\log \left(\frac{-s}{(4\pi)^2}\right)-2\right )+\{s\leftrightarrow t,u\}\right )\mathcal{M}_{\text{tree},4}\;,
\end{align}
where $\{s\leftrightarrow t,u\}$ indicates two contributions equal to the $s-$dependent one, but evaluated in $t$ and $u$ respectively. This result makes clear that the $\log s \log t$ contributions found in Eq.~\eqref{eq:singularities} are not captured by kinematic-independent IR divergences. 
Therefore, such contributions cannot be eliminated by re-dressing the one-particle states, or by redefining the asymptotic Hamiltonian that evolves the single-particle states.
Rather, the $t\to0$ singularities of Eq.~\eqref{eq:singularities} signal the IR kinematic dependence of gravitational scattering as provided by the Coulomb interaction.

%%%%%%%%%%%%%%%%%%%%%%%%%%%%%%
\section{IR effects in Dispersion Relations}\label{sec:IRE}

Dispersion relations exploit the analytic properties of amplitudes in complex energy to relate integrals of the amplitude in the IR to integrals in the UV, using Cauchy's theorem. Unitarity then implies positivity properties for the UV integrals, which translates into consistency conditions for the low energy EFT.

The IR loop effects that we have computed in section \ref{sec:loopsIR} define the analytic structure in the IR, and therefore also contribute to the IR integrals.
In this section we discuss how these dispersion relations are affected by the IR loops.

We consider two different approaches,
 dispersion relations at fixed-$t$ (FT), as developed in Ref.~\cite{Beadle:2024hqg}, and crossing symmetric dispersion relations (CS)~\cite{Sinha:2020win,Chowdhury:2022obp,Li:2023qzs,Song:2023quv,Berman:2024kdh,Zahed:2022wuy,Chowdhury:2021ynh}.
We assume that the amplitude is analytic in both $s$ and $t$ up to the physical cuts -- maximal analyticity. 
Then dispersion relations can be developed on any hyper-slice of the $s,t$ complex planes: FT and CS dispersion relations make different choices about this slices.

For clarity, we present the results specifically in $d=6$, and we  set the renormalisation scale $\mu=M$, so that our results will involve (running) Wilson coefficients evaluated at that scale.

\subsection{Fixed-$t$ dispersion relations}
\begin{figure}[t]
\begin{center}
\includegraphics[width=.6\linewidth]{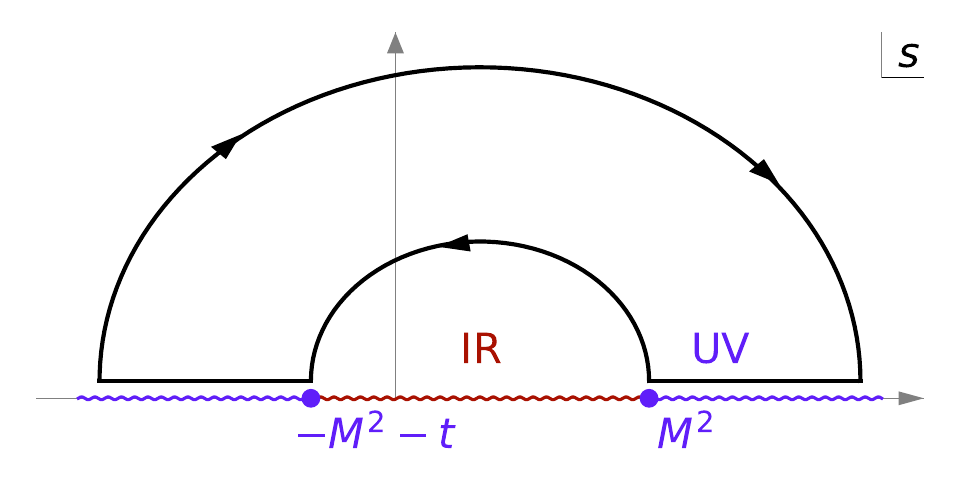}
\end{center}
\caption{\footnotesize{\emph{The analyticity structure in the complex $s\in\mathbb{C}$ plane for fixed-t amplitudes. The integral along the semicircle at infinity vanishes, implying that the IR contour integral is equal and opposite the integral along the UV discontinuity. }}}\label{fig:FTan}
\end{figure}
For fixed values of $t<0$, the discontinuity associated with  physical scattering implies the existence of a branch-cut along the entire real axis in the $s$ plane, but the amplitude is analytic elsewhere (see Section \ref{sec:loopsIR}). 
For $n\geq0$, we define arcs in their \emph{IR representation} as integrals in $s$ that probe the theory at finite energy $\bs$ and momentum transfer $q^2=-t$, and are suited  to study amplitudes with the discontinuities associated to massless particles,
\begin{equation}\label{archdeft}
\textrm{IR:}\quad\quad\quad a^{FT}_n(t)\equiv \int_{\doa}\frac{d s}{2\pi i s}\mathcal{K}^{FT}_n(s,t){\Mh(s,t)}\,,
\end{equation}
where the integral is performed along the contour $\doa$: the circle with radius $\bs +t/2$ and centred at~$-t/2$ (minus its interception with the real axis), see Fig.~\ref{fig:FTan}.
The idea is now to  exploit $s$-analyticity of the amplitude to deform the contour $\doa$ into a contour that encompasses the discontinuities on positive $s\geq \bs$ and negative $s\leq -\bs-t $ real axis, together with the semicircle at infinity.
Then, IR-UV relations follow if the kernel $\mathcal{K}_n^{FT}$  satisfies a number of conditions: \emph{i)} it has poles in $|s|\leq M^2$ such that the IR arc is non-trivial  even for analytic amplitudes, \emph{ii)} it is $s$,$u$--symmetric, allowing to easily combine the $s$ and $u$ non-analyticities and  \emph{iii)} it must decrease sufficiently fast at $s\to \infty$, $|\mathcal{K}_n^{FT}|\leq 1/s^2$. The last condition allows us to neglect the integration contour at complex infinity for amplitudes that satisfy $\lim_{|s|\to \infty} \M(s,t)/s^2 =0 $. 
For gapped theories, this  is a result of unitarity and it is guaranteed by the Froissart-Martin bound~\cite{Froissart:1961ux,Martin:1962rt,Jin:1964zza}. 
For gravitational theories a similar result applies to dispersion relations when smeared over compact impact parameter~\cite{Caron-Huot:2021rmr,Haring:2022cyf}, which is also required for convergence of the partial-wave expansion in gravity, as discussed in section~\ref{sec:loopsIR}. 
Kernels that satisfy all of these conditions can be built from
\begin{equation}\label{kernFT}
    \mathcal{K}^{FT}_n(s,t)=\frac{1}{[s(s+t)]^{n+1}} \, .
\end{equation}
Crossing symmetry in $s-u$ leads to $\Mh(s,t) =  \Mh(-s-t,t)$ -- it is also manifest in the denominator with subtractions in $s=-t$ and $s= 0$. Together with  real-analyticity $\Mh(s,t) = \Mh^*(s {}^*,t)$ it allow us to relate the integrals along the positive and negative real axis and write,
\begin{eqnarray}
 a_n^{FT}(t)
&=&\frac{1}{\pi}\int_{\bs}^\infty \frac{d s}{ s} \left(2 s+t\right)\mathcal{K}^{FT}_n( s,t)\,\textrm{Im} \M(s,t)\,.
\label{archdeftIR}
\end{eqnarray}
\begin{comment}
We use the partial-wave expansion of the amplitude,
 \begin{gather}
 \Mh(s,t)= \sum_{\ell=0}^\infty n_\ell^{(d)}\, {\cal P}_\ell\left(1+\frac{2t}{s}\right) f_\ell(s)\,,
 \end{gather}
where 
\begin{equation}
   n_\ell^{(d)} = \frac{(4\pi)^{\frac{d}{2}}(d+2\ell-3)\Gamma(d+\ell-3)}{\pi \, \Gamma\left(\frac{d-2}{2} \right)\Gamma(\ell+1)} \,,
\end{equation} 
 ${\cal P}_\ell(x)=\,_2F_1\left(-\ell, \ell + d - 3, (d - 2)/2, (1 -x)/2\right)$ are Gegenbauer polynomials. Notice that $\ell$ runs over even integers for identical scalars and the $f_\ell(s)$ coefficients are given by \eq{eq:PWs}.
 \end{comment}
The partial wave projection \eq{Eq:partial_wave_expansion} allows us to rewrite arcs in a \emph{UV representation}
\begin{equation}\label{archdeftUV}
\begin{split}
\textrm{UV:}\quad\quad\quad a_n^{FT}(t)
&=\frac{1}{\pi}\int_{\bs}^\infty \frac{ds}{s} \left(2 s+t\right)\,\sum_{\ell=0}^\infty \, n_\ell^{(d)}\,\textrm{Im}  f_\ell(s)\, \mathcal{K}^{FT}_n(s,t)\,{\cal P}_\ell\left(1+\frac{2t}{s}\right)\,\\
&\equiv\left\langle {\left(2 s+t\right)}\,\mathcal{K}^{FT}_n(s,t)\,{\cal P}_\ell\left(1+\frac{2t}{s}\right) \right\rangle_{FT} \,.
\end{split}
\end{equation}
Positivity of the integration measure $\textrm{Im}  f_\ell(s)\geq 0$ implied by S-Matrix unitarity,
 leads to a number of constraints on the UV representation of arcs, which will be interpreted as consistency conditions on the calculable IR representation of arcs.

 These constraints can be extracted by smearing both arcs in  the UV \eq{archdeftUV} and in the~IR \eq{archdeft} with appropriate functions $f(q)$ of $q^2=-t$. 
 Functions that evaluate to a positive value in the UV imply positivity conditions in the IR~\cite{Caron-Huot:2021rmr},
\begin{align}\label{smeardingexp}
     \int_0^{q_{max}} dq \,f(q) \left(2s-q^2\right)&\mathcal{K}^{FT}_n(s,-q^2)\, {\cal P}_\ell\left(1-\frac{2q^2}{s}\right) >0\quad \forall s,\ell \quad \\&\Rightarrow \quad  \int_0^{q_{max}} dq f(q) \left. a_n^{FT}(-q^2)\right|_{IR} >0\,.\nn
 \end{align}
 For such smearing functions $f(q)$ to give useful results, they must integrate to a finite quantity in the IR too, in particular on the Coulomb pole, requiring that for small $q$, $f(q)\sim q^{1+\delta}$, with $\delta >0$. On the other hand, as nicely remarked in  Ref.~\cite{Caron-Huot:2021rmr}, at large $\ell$ and large $s$ (but fixed impact parameter $b=2\ell/\sqrt{s}$) we have ${\cal P}_\ell \to \Gamma(d/2-1)J_{d/2-2}(b p)/(bp/2)^{d/2-2}$, with $J_{d/2-2}$ the Bessel function. In this limit, therefore, the UV part of \eq{smeardingexp} becomes proportional to the $d-2$ dimensional Fourier transform of $F(q)=f(q)/q^{d-3}$.
 
Now, Bochner's theorem~\cite{Rudin1990-to} requires  that functions $F$ with positive Fourier transforms must be such that the matrix $b_{ij}\equiv F(q_i-q_j)$, for all $q_{i,j}\in [0,q_{max}]$ be positive definite. Taking only two values $q_i=0$ and $q_j=q$, this condition implies that $|F(q)|\leq |F(0)|$, which translates into $4-d+\delta\leq 0$. 
This is incompatible in $d=4$ with the positivity of $\delta$ required by the Coulomb singularity, but provides a necessary condition to build the functions $f$ in higher dimensions.

\subsubsection{IR arcs}

The IR representation of arcs instead can be computed within the EFT via \eq{archdeft}, and then confronted with the UV bounds. 
At tree level the EFT amplitude is analytic when $s<M^2$ and, along with the kernels of \eq{kernFT}, finding the IR arc reduces to computing the sum of residues at $s=0$ and $s=-t$.
From Eqs.~(\ref{amptreeXY}) and (\ref{eqap:GravTree}) we  find the arcs~\cite{Beadle:2024hqg},
\begin{align}
   \textrm{Tree-level:}\quad\quad\quad a_n^{FT} (t)
        &=(\textrm{Res}|_{s=0}+\textrm{Res}|_{s=-t})\frac{\Mh(s,t)}{s[s(s+t)]^{n+1}}\nn\\
        &= -\frac{\kappa^2}{t}\delta_{n,0} +\sum_{p\geq1}\sum_{q\geq0}g_{2p+q,q}(-t)^{2(p-n-1)+q}
           \binom{p-q}{n-q}\,.
           \label{eqIRarcstree}
\end{align}
The gravity pole appears only in the first arc,\footnote{This is due to the graviton having spin-2, which forces the residue of the $t-$channel pole in the amplitude to be $s^2$, regardless of which interactions beyond minimal coupling are considered.}
$a_0^{FT}(t)=-\frac{\kappa^2}{t}+\sum_{n=1}^\infty[ n t^{2n-2}g_{2n,0}- t^{2n-1}g_{2n+1,1}]$, 
while all higher arcs are infinite polynomials in $t$.

The fact that infinitely many terms appear in \eq{eqIRarcstree} prevents any bound from being obtained through the smearing procedure in \eq{smeardingexp}, see \cite{Beadle:2024hqg,Caron-Huot:2021rmr}.
Nevertheless, it is possible to design linear combinations of dispersion relations which lead to finitely many terms in the IR. 
Ref.~\cite{Caron-Huot:2021rmr} achieves this by combining arcs $a_n$ (with $n\geq 1$) and their first derivatives evaluated at $t=0$.
Instead, in Ref.~\cite{Beadle:2024hqg} we proposed a combination,
\begin{equation}\label{eq:geralimp} 
    a_0^\text{imp}(t) =
    \sum_{n}^N\sum_k^n c_{n,k}\frac{t^{2n+k}}{k!}\partial_t^k a_n^{FT}(t)\,, 
\end{equation}
with $c_{n,k}= \frac{\partial_x^n \, \partial_y^k}{n!k!} \, G(x,y)_{x \, = \, y \, = \, 0}$ obtained from the generating function,
\begin{equation} \label{eq:impossible} 
    G(x,y)\equiv
    \frac{x\left(1+\sqrt{1-4x}-6x\right)}{\sqrt{1-4x}\left[2x+y\left(2x-1+\sqrt{1-4x}\right)\right]}
    \,.
\end{equation}
The presence of derivatives in \eq{eq:geralimp} is not in contrast with the definition of amplitudes in the distributional sense. 
Indeed, as long as the smearing measure $f(q)$ in \eq{smeardingexp} is sufficiently smooth, we can remove the derivatives by integration-by-parts leading to a linear combination of arcs weighted by different functions. 

In the IR, for $N\to \infty$, \eq{eq:geralimp} leads to $
    a_0^\text{imp}(t) =-\frac{\kappa^2}{t}+
    g_{2,0}-g_{3,1}t$, at tree-level -- in practice we will use finite $N$ in the numerics. 
Taking the same combination in the UV, bounds via smearing can now be obtained. 
As discussed in Ref.~\cite{Beadle:2024hqg}, an important aspect of \eq{eq:impossible}  is that it contains an intrinsic upper limit on 
\begin{equation}\label{eq:tmaxUB}
    |t|\leq t_{\rm max}\equiv  M^2\frac{ (\sqrt{17}-1)}{8}\,,    
\end{equation}
which limits the range on which dispersion relations can be smeared, affecting the bounds.

\subsubsection{Loop effects in IR arcs}
\label{sec:loopft}
It is at this step that the IR effects computed above play a r\^{o}le.
The different pieces in the amplitude in \eq{amp-full} give different contributions to the arcs.
The computation of the IR arcs and their derivatives, even when containing the loops, is conceptually straightforward, but the expressions eventually obtained tend to be rather complex.
Using \eq{archdeft} we compute the contour integral explicitly by introducing an angular variable $\theta$ such that,
$$
    s = -\frac{t}{2} + \left( M^2 + \frac{t}{2} \right) e^{i\theta} \, ,
$$
which is integrated from $0$ to $\pi$ and from $\pi$ to $2\pi$.
For simplicity we perform a series expansion  $|t/M^2|\ll 1$ when integrating against these kernels, justified in particular by the upper bound \eq{eq:tmaxUB} which implies $t\lesssim 0.39 \, M^2$.

\begin{figure}[h!]
    \centering
    \includegraphics[width=0.8\linewidth]{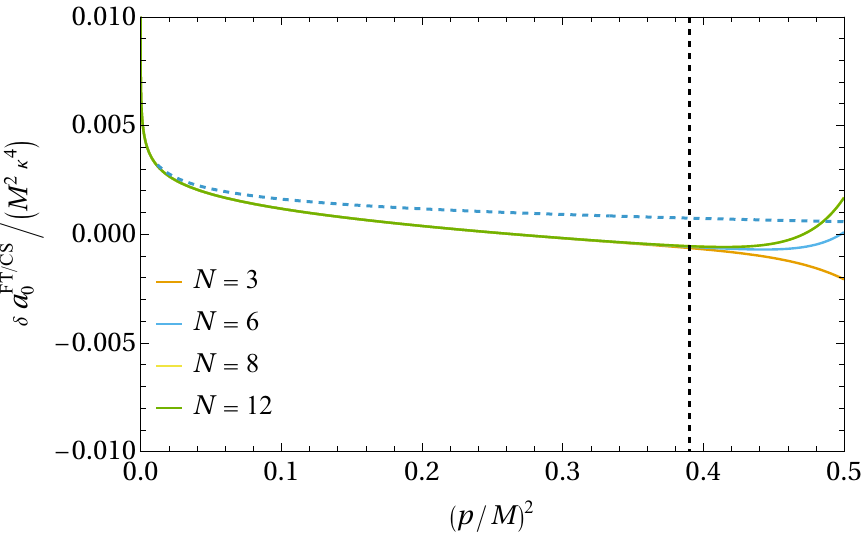}
    \caption{{\it Solid lines: 1-loop contributions to the fixed-$t$ arc $a_0^{\text{FT}}$  from the $O(\kappa^4)$ terms, for various values of $N$ in \eq{eq:geralimp}, in $d=6$. The dashed vertical line shows the radius of convergence of our expresions~\eq{eq:tmaxUB}. The dashed blue line shows the same contribution to the crossing-symmetric arc $a_0^{\text{CS}}$, where we identify $-t=p^2$. Despite the discrepancy   at large $p$, the two methods give bounds on Wilson coefficients that are in  agreement with eachother, see Fig. \ref{fig:gsandman}}}
    \label{fig:singImpArcFT}
\end{figure}

The leading terms for the first three arcs in $d=6$ dimensions (which we assume in most of the following) are,
\begin{eqnarray}
    \delta a_0^{FT} &=&\kappa^4 \Bigg(\frac{\left(49000 \pi ^2-310049\right) t+70 \log (-t) (1829 t+1050 t \log (-t)+840)+90020}{1881600 \pi ^3}\Bigg)\nonumber\\ &&
+ g_{2,0} \kappa^2 \left( -M^2 \frac{83}{4480 \pi^3} + t \frac{29}{640 \pi^3}\right) + \dots  ,\\
 \delta a_1^{FT} &  =& \nn \kappa^4 \left(\frac{-653 t+420 (7 t-4) \log (-t)+788}{53760 \pi ^3}\right) \\
    &&+ g_{2,0} \kappa^2 \left( M^2\frac{56017}{470400 \pi^3} - t \frac{369}{4480 \pi^3} \right) + \dots \, ,\\
      \delta a_2^{FT} &=& \nn \kappa^4 \left(\frac{477 t+210 (27 t-8) \log (-t)-332}{161280 \pi ^3}\right) \\
        &&+ g_{2,0} \kappa^2 \left(M^2 \frac{83}{4480 \pi^3} - t\frac{107}{2688 \pi^3} \right) + \dots 
    \end{eqnarray}
where the dots contain contributions from all combinations of EFT coefficients as well as  higher powers in $t$. We show these results, for the gravity-only contribution, in Figure~\ref{fig:singImpArcFT}.

We see that any discontinuity in $s$ induced by loop effects in the amplitude, appears in \emph{every} arc $a_n^{FT}(t)$. 
So, contrary to the tree-level idealisation where the $t$-pole appears only in the first arc, here the non-analytic functions  like $\log(-t/s)$ will propagate to all arcs.
The most important such pieces are associated with the gravitational interaction only. In the $|t|\ll M^2$ limit these are easy to identify in any dimension, and contribute to the arcs as,
\begin{equation}\label{eq:behaviorGravLoopArc}
    \delta a_n^{FT} \underset{|t| \ll M}{\sim} \frac{\kappa^4M^{d-8n-2}t^{\frac{d - 6}{2}}}{2n-1}\begin{cases} \log \left( -t/M^2 \right) & d \text{ is even,} \\ 1 & d \text{ is odd} \,.\end{cases}
\end{equation}

\subsection{Crossing Symmetric Dispersion Relations}\label{sec:CSDR}
Naive fixed-$t$ dispersion relations have the inconvenience that already at tree-level, they contain infinitely many EFT coefficients in the IR, as discussed below \eq{eqIRarcstree}.
The reason behind the appearance of infinitely many coefficients is that dispersion relations correspond at tree-level to $n$-residues in $su=0$; these  are not aligned with the crossing-symmetric expansion of the amplitude \eq{amptreeXY}, where $(stu)^m$ and infinitely many terms $(s^2+t^2+u^2)^n=2^n(su+st+tu)^n$ contain the same powers of $su$. 

For this \emph{not} to be the case, only $(stu)^m$ and $(s^2+t^2+u^2)^m$ must  appear in the same dispersion relation, and nowhere else.\footnote{Similarly, it might also be possible to obtain finitely many terms if  $(stu)^m$ and $(s^2+t^2+u^2)^{m'}$, with $m\neq m'$ appear in the same dispersion relation.}
The simplest way to realise this is  to choose, instead of $s$ and $t$, new coordinates  with the property
\begin{equation}\label{eq:csdrdef}
    \frac{2 stu}{(s^2+t^2+u^2)}=p^2
\end{equation}
where $p^2>0$ is held fixed in dispersion relations.
Crossing symmetric dispersion relations are developed along variables with the property \eq{eq:csdrdef}. 

\begin{figure}[H]
    \centering
\includegraphics[height=0.367\linewidth,valign=t]{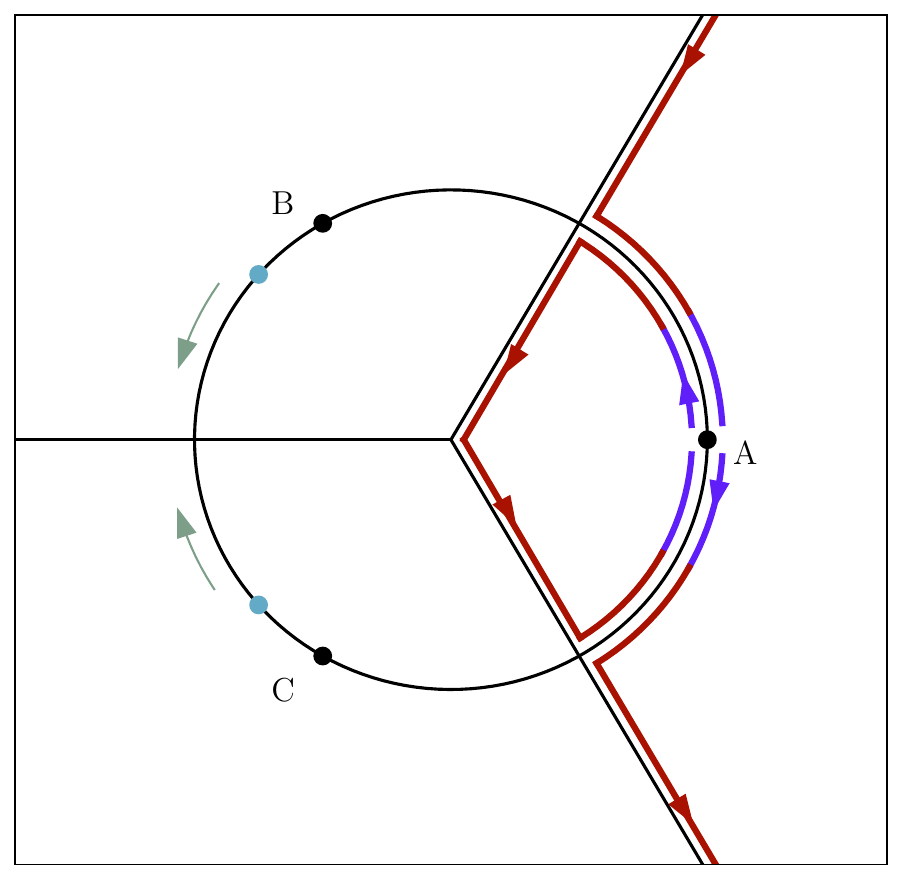}
\hspace{1cm}
\includegraphics[width=0.4\linewidth,height=0.4\linewidth,valign=t]{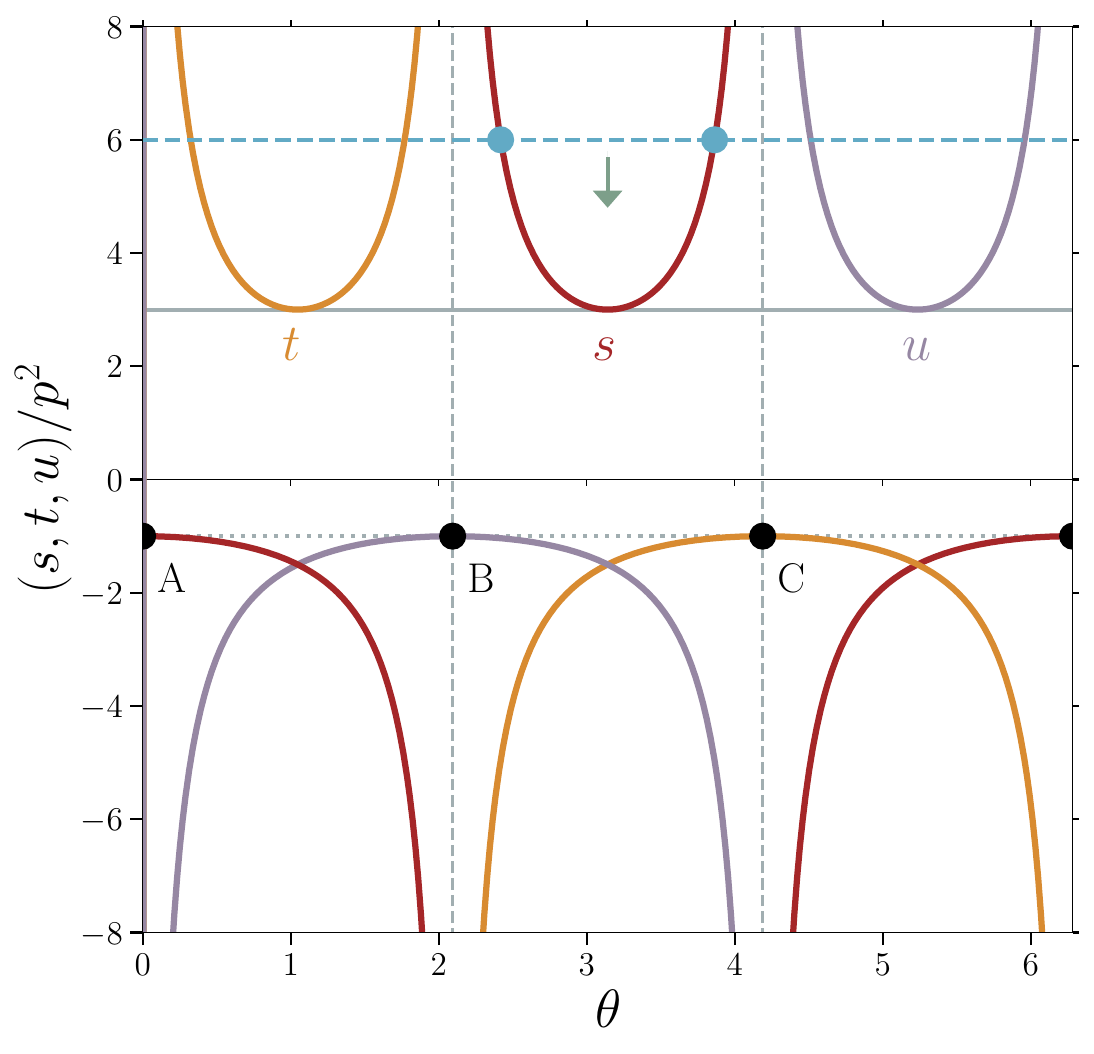}
    \caption{{\it LEFT: Analytic structure of the amplitude in the $z\in\mathbb{C}$ plane, with branchcuts on the unit circle and the radial directions $-1,-\xi,-\xi^2$. In blue(red) the UV(IR) contours of integration $C_{\rm UV}$ ($C_{\rm IR}$) from~\eq{uvircontours} (which include the analog around the points B and C). On the unit circle the amplitude is real; points and arrows help translating to the right panel.
    RIGHT: Values of $s,t,u$ along the unit circle as a function of the angle $\theta$. At the black points A,B,C two of the Mandlestam variables diverge, with the third asymptoting to $-p^2$. On the positive side we have $s,t,u>3p^2$ and on the negative $s,t,u<-p^2$. In blue, points of fixed $s=M^2>3p^2$; as $M$ is lowered the points move closer to the radial directions, as shown by the arrows.}}
    \label{fig:analyticCSDR}
\end{figure}

For maximally analytic amplitudes -- analytic in both $s$ and $t$ -- dispersion relations can be developed on any slice of the $s,t$ complex planes. 
In particular the slice of constant $p$ in \eq{eq:csdrdef} is what we are interested in.
Following \cite{Li:2023qzs}, we change variables from $(s,t)$ to $(z,p)$,
\begin{equation}
    s(z,p)=-\frac{3p^2 z}{1+z+z^2}\,,\quad t(z,p)=s(z\, \xi,p)\,,\quad u(z,p)=s(z \,\xi^2,p)\,,
\end{equation}
with $\xi=e^{i 2 \pi /3}$ and $0<p^2$ satisfying \eq{eq:csdrdef}. 
The amplitude is analytic in $z\in \mathbb{C}$, up to the physical cuts corresponding to real positive values of $s,t,u$. 
These are located on the unit circle, where all Mandelstam variables are real (e.g. $s = -3p^2 / [ 1 + 2 \cos \theta ]$), and on the lines that span from the origin in the directions $-1,-\xi,-\xi^2$, see the left panel of~Fig.~\ref{fig:analyticCSDR}. 

As illustrated in the right panel of Fig.~\ref{fig:analyticCSDR}, the points
 $z=1,\xi,\xi^2$ correspond to infinities in one of the (real) Mandelstam variables; for instance approaching $z=\xi^2$ from above corresponds to $s\to \infty$, $t=-p^2$. 
 We build dispersion relations starting from knowledge of the amplitude's behavior in these limits. Along the discussion below \eq{archdeft}, we assume that   amplitudes {\it smeared in $p$} grow slower than $s^2$ at large $|s|$,
\begin{equation}\label{froissy}
     \lim_{s\to \infty}\int_0^{p_{max}} dp f(p) \frac{\mathcal{M}(z,p^2)}{s^2}=0\,. 
 \end{equation}
where $\mathcal{M}(z,p^2)=\mathcal{M}(s(z,p),t(z,p))$\,.
From \eq{froissy} we can write,
\begin{equation}\label{vanDR}
\oint_{z=1,\xi,\xi^2} \frac{dz}{4\pi i}\, \mathcal{K}^{CS}_n(z) \mathcal{M}(z,p^2)\equiv 0\,,
\end{equation}
where it is implicitly assumed that these relations will be smeared in $p$. 
The kernel $\mathcal{K}^{CS}_n$ is built according to the following criteria: it is invariant under $z\rightarrow \xi z$ and $z\rightarrow \xi^2 z$, as imposed by crossing symmetry. 
Then, in the Regge limits $z=(1,\xi^2, \xi^{-2})$ it contains enough subtractions to ensure convergence as assumed in Eq. \eqref{froissy}, which is explicitly realised by a $(1-z^3)^{2n+1}$ pole.
Finally, the kernel contains a low energy symmetric pole at $z=0$, i.e. $s=t=u=0$, in order to capture the EFT contributions.  
We then consider the kernel
\begin{equation}
\mathcal{K}^{CS}_n(z)= (-1)^{n} \frac{\left(z^3+1\right)\left(1-z^3\right)^{2n+1}}{3^{3(n+1)}p^{4 n + 4}z^{3n+4}}\,,
\end{equation}
with $n\geq 0$.

The contour in \eq{vanDR} can be deformed and separated into two (equivalent) pieces, as illustrated in the left panel of Fig. \ref{fig:analyticCSDR}, where the separation is defined  by the value  $s=M^2$, $t=M^2$ and $u=M^2$ on the unit circle (see green dots in left and right panels),
\begin{equation}\label{uvircontours}
a_n^{CS}(p)\equiv\oint_{C_{\rm IR}}\frac{dz}{4\pi i}\, \mathcal{K}^{CS}_n(z)\mathcal{M}(z,p^2)=-\oint_{C_{\rm UV}}\frac{dz}{4\pi i}\, \mathcal{K}^{CS}_n(z)\mathcal{M}(z,p^2)\,.
\end{equation}
Now the UV representation of $a_n^{CS}$ is equivalent to the integral along the (positive) amplitude discontinuity, and can be rewritten in terms of a more familiar integral in $s$,
\begin{equation}
    a_n^{CS}(p)=\int_{M^2}^{\infty}\frac{ds}{2\pi}s^{-{3 n}-4} \left(3 p^2+2 s\right) \left(p^2+s\right)^{{n}} {\rm Disc} \mathcal{M}(s,p^2)\,.\label{UVCSDR}
\end{equation}
where $t$ (and $u$) read,
\begin{equation}\label{eq:tuasps}
t=-\frac{s(p^2+s-\sqrt{s-3p^2}\sqrt{s+p^2})}{2(s+p^2)}\,,\quad u=-\frac{s(p^2+s+\sqrt{s-3p^2}\sqrt{s+p^2})}{2(s+p^2)}\,.
\end{equation}
So, in contrast to fixed-$t$ dispersion relation here $t$ changes with $s$: for $s\to \infty$ we have $t\to-p^2$, while for $s\to 3p^2$ (the minimum value on the unit circle) $t\to-3p^2/2$.
Moreover, since we need $M^2\geq 3p^2$ (see the dashed line in the left panel of Fig.~\ref{fig:analyticCSDR}), this implies that 
$
    |t|\leq\frac{M^2}{2}\,,
$
while for values larger than this the role of $t$ is taken by $u$, as shown in the figure.
This implies that CS dispersion relations have a naturally built-in upper bound on the allowed smearing range 
\begin{equation}
    p^2\leq \frac{M^2}{3}\,,
\end{equation}
which translates into $t_{max}=\frac{M^2}{2}$.

In the tree-level approximation, the IR contour has no non-analyticities and the only contributions come from poles at $z=0$ and $z=\infty$.
Instead, in realistic amplitudes like the ones computed in section \ref{sec:loops}, the IR contour involves a complicated sum over non-analyticities in $z$ across the unit circle and the radial directions. 

We circumvent this problem with a trick.
The EFT series expansion  converges in the IR $s\lesssim M^2$ where the amplitude can be well approximated by finitely many terms, including the tree-level polynomial part and the non-polynomial loop contributions described above.
The {\it truncated function} has non-analyticities associated only with the discontinuities of the loop contributions and the subtraction.
These are  known a priori in the whole complex plane and they consist of the regions already illustrated in Fig.~\ref{fig:analyticCSDR} but also include {\it extra} poles in the points $z\in\mathcal{D}=\{1,\xi,\xi^2\}$. 
We will use this analytic continuation to compute the IR arcs along the UV contours plus the poles in $\mathcal{D}$. 
Importantly, this is different from the actual untruncated UV amplitude which receives further contributions to the discontinuities from UV states and has vanishing residues at infinity. 

\begin{figure}
    \centering
\includegraphics[width=0.4\linewidth]{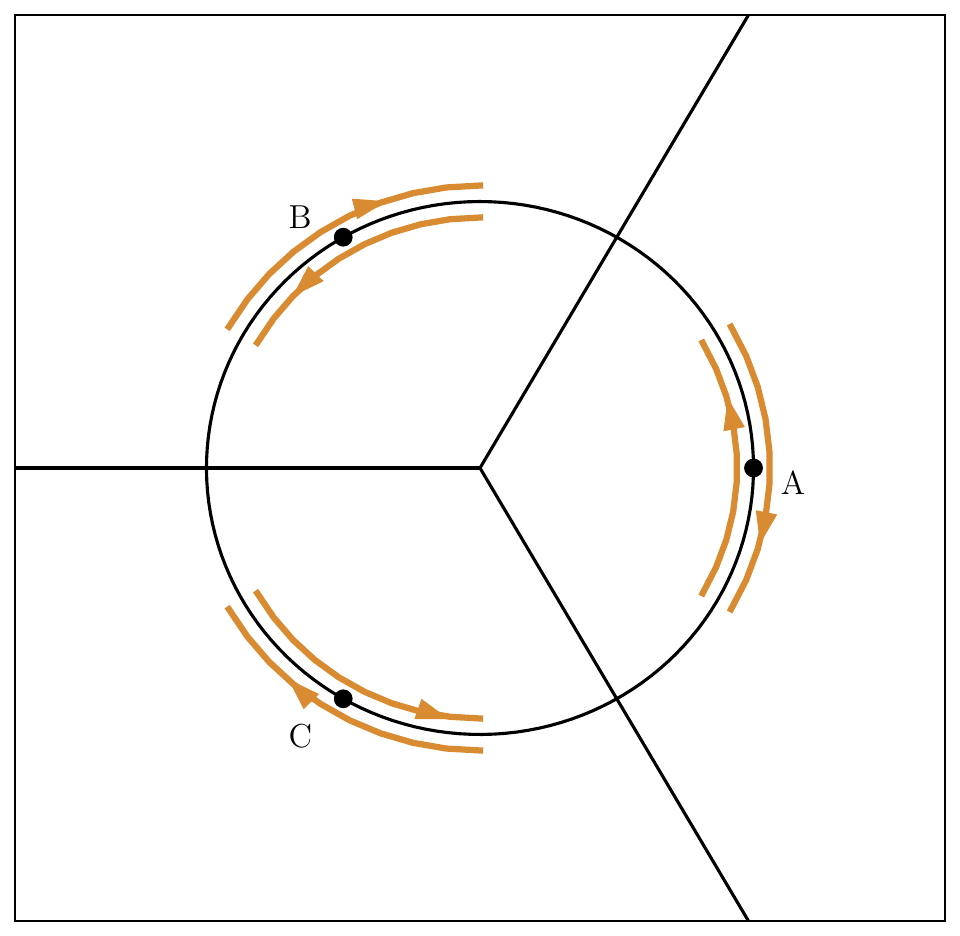}
    \caption{{\it Alternative  contour of integration for the IR EFT amplitude. The contour includes the integrals along the discontinuity and the poles around A, B  and C, both of which separately diverge. We regularise them by integrating at a finite distance from the poles and the singularities cancel out.}}
\label{fig:analyticCSDRIR}
\end{figure}

Therefore the IR integrals can be written as,
\begin{equation}
    a_n^{CS}(p)\equiv\oint_{C_{\rm UV+\mathcal{D}}}\frac{dz}{4\pi i}\, \mathcal{K}_n^{CS}(z)\mathcal{M}^{\text{EFT}}(z,p^2)
\end{equation}
as illustrated in Fig.~\ref{fig:analyticCSDRIR}. 
As it is clear from \eq{UVCSDR}, the integral of the EFT amplitude along the UV contour diverges -- for instance a $s \log(-s) $ term has a $\pi s$ discontinuity that diverges quadratically. However, this divergence is cancelled by an equal and opposite contribution from the poles at $z=1,\xi,\xi^2$, leaving only finite pieces, as it should. With this method we compute below the IR part of CS dispersion relations, see section \ref{sec:loopscsdr}.

Meanwhile, the bounds will come from the UV representation, after writing it in terms of partial waves via \eq{Eq:partial_wave_expansion},
\begin{equation}
    a_n^{CS}(p)= \sum_{\ell=0}^\infty n_\ell^{(d)}\int_{M^2}^{\infty}\frac{ds}{2\pi}\frac{{\text Im} f_\ell(s)}{s^{{3 n}+4}} \left(3 p^2+2 s\right) \left(p^2+s\right)^{n}  \,\mathcal{P}_{\ell}\left(\sqrt{\frac{s-3p^2}{s+p^2}}\right),
\end{equation}
which can also be expressed in the form 
$
    a_n^{CS}(p)=\left\langle I_{\ell,n}^{CS}(s, p) \right\rangle_{CS}
$, 
with the definitions, 
\begin{equation}
    \langle\dots\rangle_{CS}= \sum_{\ell=0}^\infty n_\ell^{(d)}\int_{M^2}^{\infty}\frac{ds}{2\pi}\frac{{\text Im} f_\ell(s)}{s^{{3 n}+4}}\left(\dots\right)\,,\quad  
    I_{\ell,n}^{CS}(s,p)=\frac{3 p^2+2 s} {\left(p^2+s\right)^{-{n}}}  \,\mathcal{P}_{\ell}\left(\sqrt{\frac{s-3p^2}{s+p^2}}\right)\,.
\end{equation}

At tree level, using the explicit form of the amplitude Eqs. (\ref{amptreeXY}) and (\ref{eqap:GravTree}), we obtain the following IR-UV relation for $n=0$,
\begin{equation}
\label{eq:firstarc_ir_uv}
    a_0^{CS}(p)= \frac{\kappa^2}{p^2}+ g_{2,0} + g_{3,1}p^2 =\left\langle (2s +3p^2) \mathcal{P}_{\ell}\left(\sqrt{\frac{s-3p^2}{s+p^2}}\right)\right\rangle_{CS}\,.
\end{equation}
Notably, on the IR part only a finite number of terms appear, as discussed above \eq{eq:csdrdef}.

We can now perform the same procedure used for the fixed-$t$ case and smear the IR and UV sides, now in $p\in [0,\,M^2/\sqrt{3}]$,

which gives (for $n=0$), 
\begin{equation}\label{eq:CSIRUV}
    \int_0^{M^2/\sqrt{3}}\dd p \, f(p) \left( \frac{\kappa^2}{p^2}+ g_{2,0} + g_{3,1}p^2 \right)=  \int_0^{M^2/\sqrt{3}}\dd p \, f(p)\left\langle I_{\ell,0}^{CS}(s,p)\right\rangle.
\end{equation}
Exploiting convergence of the partial wave expansion in the UV and finiteness of the dispersion relations, we  swap the order of integrations $\dd s$ and $\dd p$. 
Then, \textit{ if} each element of the UV sum in $\ell$ is positive for each value of $s$ after being integrated in $\dd p$, \textit{then}  the integral on the IR side is positive,
\begin{equation}\label{eq:csdr_bound_procedure}
    \int_{0}^{M^2/\sqrt{3}}\dd p \;f(p)\;I_{\ell,n}^{CS}(s,p)>0, \; \forall s \;{\rm and}\; \forall\; \ell \implies \int_0^{M^2/\sqrt{3}}\dd p\;  f(p)\; a_{n,{\text IR}}^{CS}>0.
\end{equation}

\subsubsection{Loop Effects in Crossing Symmetric Dispersion Relations}\label{sec:loopscsdr}
Following the approach described in Fig.~\ref{fig:analyticCSDRIR}, we compute the contribution of loop effects on the IR CS dispersion relations.
The leading EFT interactions give,
\begin{eqnarray}
   \delta a^{CS}_0 &=&-\frac{g_{2,0}^2}{107520 \pi^3} \left[ 60 + 133 \Bar{p}^2 - 2 \left( \Bar{p}^2 \right)^2 + 2 \left( \Bar{p}^2 \right)^3 \log \left( 1 + \frac{1}{\Bar{p}^2} \right) \right] \nn\\
    &&- \frac{g_{2,0} g_{3,1}}{161280 \pi^3} \left[ 33 + 70 \Bar{p}^2 + 3 \left( \Bar{p}^2 \right)^2  - 6 \left( \Bar{p}^2 \right)^3 + 6 \left( \Bar{p}^2 \right)^4 \log \left( 1 + \frac{1}{\Bar{p}^2} \right) \right]\nn\\
    &&+ \frac{g_{2,0} g_{4,0} }{4838400 \pi^3} \left[ -2616-4755 \Bar{p}^2 + 100 \left( \Bar{p}^2 \right)^2 -150 \left( \Bar{p}^2 \right)^3 + 300 \left( \Bar{p}^2 \right)^4 - 300 \left( \Bar{p}^2 \right)^5 \log \left( 1 + \frac{1}{\Bar{p}^2} \right) \right]\nn\\
    &&+ \frac{g_{3,1}^2}{3225600 \pi^3} \left[  -72 -105 \Bar{p}^2 + 20 \left( \Bar{p}^2 \right)^2 - 30 \left( \Bar{p}^2 \right)^3 + 60 \left( \Bar{p}^2 \right)^4 - 60 \left( \Bar{p}^2 \right)^5 \log \left(1 + \frac{1}{\Bar{p}^2} \right) \right] \nn\\
    &&+ \dots\,,
\end{eqnarray}
while mixed EFT-gravity effects give,
\begin{eqnarray}
   \delta a^{CS}_0 &=&
    \kappa^2 \Bigg[\frac{g_{2,0}}{4480 \pi^3} \left( 83 + 203 \Bar{p}^2 - 23 \left(\Bar{p}^2\right)^2 \log \left( 1 + \frac{1}{\Bar{p}^2} \right) \right) \nn\\
   &&+  \frac{g_{3,1}}{4480 \pi^3} \left( 10 + \frac{91}{2} \Bar{p}^2 + 23 \left( \Bar{p}^2\right)^2 - 23 \left( \Bar{p}^2 \right)^3 \log \left( 1 + \frac{1}{\Bar{p}^2} \right) \right)\nn\\
   & &+ \frac{g_{4,0}}{241920 \pi^3} \, \left( \frac{1914 + 4250 \Bar{p}^2  + 1721 \left( \Bar{p}^2 \right)^2 + 1119 \left( \Bar{p}^2 \right)^3 + 2238 \left( \Bar{p}^2 \right)^4}{1+ \Bar{p}^2} \right.\nn \\
   & &\left. - 2238 \left( \Bar{p}^2 \right)^4 \log \left( 1 + \frac{1}{\Bar{p}^2} \right)  \right) +\dots \Bigg]\,.
\end{eqnarray}
where we define $\Bar{p} = p / M$.
In pure gravity, we instead obtain a long expression which we report in Appendix \ref{app:arcs}.
Figure~\ref{fig:singImpArcFT} provides a graphical representation of these results, and also compares it with the FT approach.

It is instructive to isolate the leading effects at small $p/M$, which can be computed using the leading discontinuity across the unit circle in $z\in \mathbb{C}$ (i.e.  real kinematics) from the box diagram in \eq{eq:singularities}.
Similarly to what found for arcs at fixed $t$ in \eq{eq:behaviorGravLoopArc}, this discontinuity will contribute to all arcs. Following \eq{UVCSDR}, we are able to find a compact expression in all dimensions,
\begin{equation}\label{eq:arcsptozero}
    \delta a_n^{CS} \underset{|t| \ll M}{\sim} \frac{\kappa^4M^{d-4(n+2)}p^{d-6}}{n }\begin{cases} \log \left(\bar{p}^2 \right) & d \text{ is even,} \\ 1 & d \text{ is odd} \,.\end{cases}
\end{equation}

\subsection{Impact for Dispersion Relations}

From the above discussion we see that IR loops affect dispersion relations in several ways.
First of all, they introduce non-analyticities, which exhibit  certain singular behaviours in dispersion relations.
These effects can be classified into two categories. The first involves contributions that grow as $t\to 0$, such as those described in \eq{eq:behaviorGravLoopArc}, or their crossing-symmetric  counterparts as $p\to 0$, see \eq{eq:arcsptozero}. These contributions impose limitations on how dispersion relations can be applied to extract bounds. Specifically, they affect dispersion relations with any number of subtractions, unlike tree-level dispersion relations where, even in the presence of gravity, non-analyticity arises only in $a_0$, as shown in \eq{eqIRarcstree}.
The gravitational effects highlighted in \eq{eq:behaviorGravLoopArc} lead to divergent  FT arcs at $t=0$ for $d\leq 5$, with their derivatives diverging for $d\leq8$. By contrast, in $d\geq 3$, EFT-only interactions result in arcs and their first derivatives that remain regular as $t\to0$, as also discussed in~\cite{Arkani-Hamed:2020blm,Bellazzini:2021oaj}. 
This remains true also for the CS case. 

This distinction shows the necessity of moving beyond the improvement procedure of Ref.~\cite{Caron-Huot:2021rmr}, which relies on higher arcs and their derivatives in the forward limit. Instead, the methodology of Ref.~\cite{Beadle:2024hqg}, which is fully defined away from $t=0$, provides a more suitable framework to address thes problems.

On the other hand, divergences in $M^2\to0$ are the reflection of the fact that the running induced by more relevant operators sooner or later dominates over that induced by the less relevant ones, as discussed in section~\ref{sec:tree} and detailed in the context of dispersion relations in Ref.~\cite{Bellazzini:2020cot}.
 As discussed in this reference, the first operators that exhibit running is crucial for the following reason: at small $M$, the bounds on the leading running operator are modified by $\log{ M}$ corrections. However, by dimensional analysis, the bounds on less relevant coefficients are influenced by terms proportional to powers of $1/M$, which are much larger at small $M$.
 In $d\geq 5$, the bounds on $\{\kappa, g_{2,0}, g_{3,1}\}$ are impacted solely by  $\log{ M}$ effects along $g_{3,1}$ as $M\to 0$. Polynomial effects would instead enter the bounds of the more irrelevant coefficients, like $g_{4,0}$ etc, which can now violate tree-level positivity and become negative by amounts that are polynomial in $1/M$~\cite{Bellazzini:2020cot}.

\vspace{2mm}
Another important aspect of loop effects is that they imply that
\emph{all} dispersion relations contain \emph{all} couplings. This stems from the fact that the discontinuity  is proportional to the entire amplitude, involving all coefficients. 
This is also in sharp contrast with the tree-level limit. There, in the vanishing coupling limit ($g,\kappa M^{d-2}\to 0$ in sec.~\ref{sec:tree}), the boundness of the EFT expansion parameter $E/M\lesssim 1$  emerges as a result of dispersive bounds, \cite{Bellazzini:2020cot,Caron-Huot:2020cmc}. At finite coupling, instead, it becomes unfeasible  to derive sharp results because arcs involve infinitely many couplings, appearing linearly and quadratically.  To extract quantitative results we will have to make \emph{a priori assumptions} about the size of the higher coefficients.

Finally, an intriguing implication of loop effects is that, as discussed at the beginning of this paragraph, they require consistent dispersion relations without forward limits. The two such examples, at fixed-$t$ \cite{Beadle:2024hqg} and in crossing-symmetric dispersion relations \cite{Sinha:2020win,Li:2023qzs}, operate within a naturally compact range in $t$. This range, determined by the smearing procedure, is restricted to at most $|t|<M^2/2$.
The size of this range will have an important effect on the bounds, even at tree-level.

\section{Bounds on Gravitational Amplitudes}\label{sec:BoundsMod}

The arcs defined in dispersion relations are non-perturbative objects. When computed in the IR at tree-level they take the simple form like \eq{eq:firstarc_ir_uv}, in terms of the Lagrangian couplings and when computed at loop-level they also involve the corrections that we have computed, like \eq{eq:csGRpiece}. 
At stronger coupling higher loops will also appear in the IR expression. 
The UV expression, on the other hand, is always the same.

To answer the question of how loop effects impact the bounds, one would like to define non-perturbative objects that in the limit of weak coupling match to $\kappa,g_{2,0},g_{3,1}$, etc. 
Unfortunately there are infinitely many such combinations, because there are infinitely many functionals $f(p)$ that integrate to the same quantity in the IR \eq{eq:CSIRUV}~\footnote{In the IR the 0-th tree-level arc contains only powers $1/p^2,p^0,p^2$: there are infinitely many functionals of the form $p^2\times Pol(p)$, with $Pol(p)$ a polynomial, which are positive in the UV, but vanish in the IR. For instance, because of orthogonality of the Legendre polynomials, any $Pol(p) =\sum _{l>4}P_l(p)$ integrates to 0 in the IR of \eq{eq:CSIRUV}.}.

Therefore here  we take the following approach. 
We first compute the \emph{tree-level} bounds on ratios of the most relevant couplings $g_{2,0}/\kappa^2$ and $g_{3,1}/\kappa^2$. 
For every point on the boundary there will be an associated functional~$f(p)$. 
We then evaluate this functional on the loop contributions to derive the modification of the bounds. In Appendix~\ref{sec:HOeff} we compare this approach to that of exploring more generic loop-level functionals, and verify that this only leads to a small change in the bounds -- see Fig.~\ref{fig:loopExplicit}.

\subsection{Tree-level Bounds }\label{sec:treelevelreuslts}
\begin{figure}
    \centering
\includegraphics[width=0.8\linewidth]{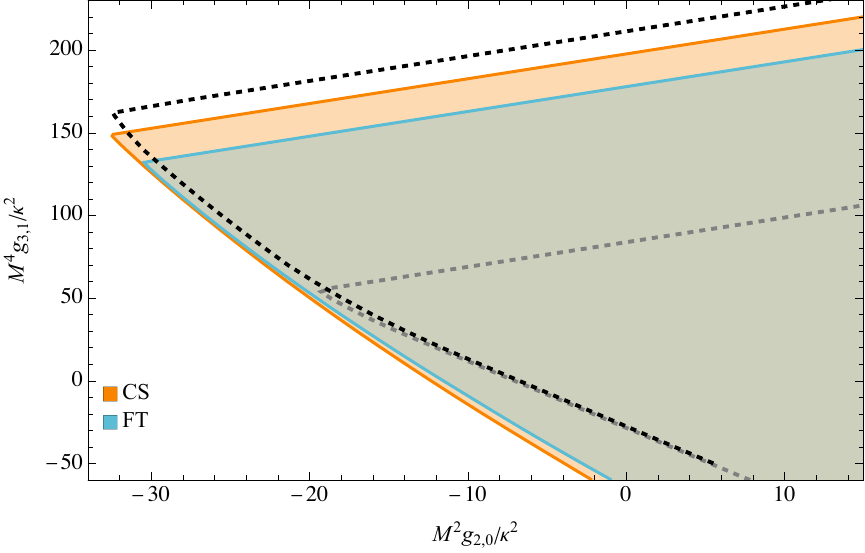}
   \caption{{\it Bounds on $ g_{2,0}$ and $g_{3,1}$ in units of $\kappa^2$, for CS dispersion relations and the FT with the improvement of Ref.~\cite{Beadle:2024hqg}, in $d=6$ dimensions. In gray (black) dashed the results from Ref. \cite{Caron-Huot:2021rmr}  with smearing  up to $t_{\text{max}}=M^2$ ($t_{\text{max}}=M^2/2$). The bound from CS dispersion relations is obtained with $\ell_\text{max}=30$, a basis of 20 functionals and 10 values of $s$ as described in Appendix \ref{app:smearing}. The bound from FT was computed with $\ell_\text{max}=400$ and a basis of 8 functionals and degree of improvement $N=12$ consistently with Ref.~\cite{Beadle:2024hqg}. }}
\label{fig:wedge_results}
\end{figure}

We derive the bounds using different methods, based on
 FT and CS dispersion relations,
as described in \eq{smeardingexp} and \eq{eq:csdr_bound_procedure}.
We begin by examining the tree-level bounds and later show how loop effects alter them.

Using only the tree-level amplitude, in Fig.~\ref{fig:wedge_results} we show the bounds on the first two EFT coefficients $g_{2,0}$ and $g_{3,1}$, normalised by the strength of gravity $\kappa^2$, in units of the EFT cutoff~$M$.
The different lines compare results obtained with CS (orange) and FT (blue) dispersion relations.
For comparison, in black (grey) the results obtained with the method of Ref.~\cite{Caron-Huot:2021rmr} for smearing in $-t_{\rm max}<t<0$ with $t_{\rm max}=M^2/2$ ($t_{\rm max}=M^2$). 
The black line has the same upper limit on $|t|$ as the CS method discussed in this work, while the grey one extends to higher values, hence explaining the tighter constraints.

 All methods exhibit the same  asymptotic behaviour, with slope
\begin{equation}\label{eq:asymptbounds}
\frac{M^2 g_{2,0}}{\kappa^2}\to \infty \quad \Rightarrow\quad   
-4.07\lesssim\frac{M^4 g_{3,1}/\kappa^2}{M^2 g_{2,0}/\kappa^2} \leq \frac{3}{2},
\end{equation}
compatible with bounds in the absence of gravity \cite{Caron-Huot:2020cmc} -- this result is  not obvious in Fig.~\ref{fig:wedge_results} and is highlighted in Fig.~\ref{fig:asymptotics} in the Appendix. 
In this limit gravity becomes negligible and the functionals can move closer to the near-forward region where \eq{eq:asymptbounds} holds.
Moreover, in all methods, the upper bound always saturates the asymptotic slope. 
This is consistent with the expectation in terms of UV models: exchange of a mass $M$ scalar provides a consistent UV completion and gives an EFT amplitude with $g_{3,1}M^2/g_{2,0}=3/2$, corresponding to the steepness of the upper bound in~Fig.~\ref{fig:wedge_results}.

We believe the small difference  in the upper bound observable in  CS and FT results 
can be traced to numerics rather than physics.
In particular, the two methods have very different convergence properties, with CS involving  heavier initial computations but converging faster, while FT needing more constraints to stabilize~\cite{Beadle:2024hqg}.
The results shown have converged within our computational abilities; we show more details of this in Fig.~\ref{fig:Jcheck} of  Appendix~\ref{app:smearing}. 
It is plausible however that including more values of $\ell$ as well as  larger bases for the functionals $f(p)$, would bring the two methods to agree.

The situation is more interesting and more complicated on the lower bound, where the  methods of Ref.~\cite{Caron-Huot:2021rmr} differ the most from the CS and FT ones.
In principle, it is not surprising that the figures appear different since they are based on different assumptions about the analytic structure in $s$ and~$t$, with Ref.~\cite{Caron-Huot:2021rmr}  extending to $t=0$ in higher subtracted dispersion relations.
The different methods  also employ different kernels,  reflecting the underlying assumptions on the amplitude, and these kernels have different behaviours.
In particular,  CS  and FT relations are limited to $|t|<M^2/2$, while the dispersion relations of Ref.~\cite{Caron-Huot:2021rmr}  can in principle extend to larger values of $|t|$. 
The range $|t|\leq M^2/2$, corresponds to arcs  limited to physical scattering angles~$\theta\leq \pi/2$ and is motivated by 
 crossing symmetry, which implies that larger values of $|t|$ would be redundant under $u\leftrightarrow t$ exchange.

 It would be interesting to develop a sharper perspective on extremal UV amplitudes, to identify what theories -- if any -- satisfy our bound, but not the one of Ref.~\cite{Caron-Huot:2021rmr}, smeared over larger values of $t$.
It was already pointed out in Ref.~\cite{Caron-Huot:2021rmr} that for theories with 
only a finite number of UV weakly coupled particles of finite $\ell$ with masses $m_\ell\leq \tilde M$,  the residues are finite polynomials in $t$ and dispersion relations hold to $|t|\leq \tilde M$.
Interestingly,  the $stu$-model proposed in Ref.~\cite{Caron-Huot:2020cmc}, with UV amplitude, 
\begin{equation}\label{eq:stu}
    {\cal M}=\frac{1}{(s-M^2)(t-M^2)(u-M^2)},
\end{equation}  
would appear to violate this, since the simultaneous poles in the $s$, $t$ and $u$ channels can be thought of as the exchange of infinitely many particles with all spins at $s=M^2$, thus implying possible non-analyticities since $P_{\ell}(1-2t/M^2)$ diverges as $\ell\to \infty$ for $|t|>M^2$. 
However, for $s=M^2$ and negative $t$, \eq{eq:stu} becomes singular only at $t=-2M^2$, when the $u$-pole is hit. 
So, even the $stu$ model, despite its accumulation point, has amplitudes that are smearable up to $|t|=2M^2$.
On the other hand,  amplitudes in gravity including loops have a smaller cutoff. 
Indeed $\mathcal{I}_{\Box}$, evaluated at $s=M^2$, is singular for $t<-M^2$, see \eq{eq:expsing} ($\mathcal{I}_{\bigcirc}$ instead depends only on one kinematic variable and does not pose any problem).

In summary, 
while tree-level amplitudes at fixed $s=M^2$ are smooth over a broad range in $t<0$, gravity at finite coupling imposes $t>-M^2$. However, the dispersion relations that remain finite with gravity loops, imply the more stringent condition  $t>-M^2/2$, possibly implied by crossing symmetry.

\vspace{2mm}

\subsubsection{Higher Couplings at tree-level}\label{sec:HOtree}
At tree-level, higher arcs  $a_n$ with $n>0$ don't have the graviton pole. Bounds between the higher couplings can therefore be derived with the simpler methods of Refs.~\cite{Arkani-Hamed:2020blm,Bellazzini:2020cot,Caron-Huot:2020cmc}. 
In particular one finds that, starting from  $g_{4,0}$,
the coefficients are monotonically decreasing in units of $M$ -- up to computable $O(1)$ numbers that depend also on the coupling normalisation. For instance, 

\begin{equation}\label{eq:monotonicity}
    0 \leq M^4\frac{g_{6,0}}{g_{4,0}} \leq 1\,,
\end{equation}
and so on, for other coefficients. Moreover, there are also bounds on $g_{4,0}$ in units of $\kappa$.

This is an important result: qualitatively, higher coefficients respect dimensional analysis. Without gravity this convergence starts already at $g_{2,0}$ (e.g. $0\leq{M^4g_{4,0}}/{g_{2,0}}\leq1$ ), but gravity deforms this statement. With gravity it is possible to have $g_{2,0}=0$ or $g_{3,1}=0$, but $g_{4,0}>0$, and then the coefficients respect monotonicity, as in \eq{eq:monotonicity}.
Indeed, this is the case for dilaton scattering in Type II String theory, where $g_{2,0}=g_{3,1}=0$, despite gravity and the other coefficients being sizeable.

\subsection{Bounds with Finite Couplings}

At finite coupling, the IR expressions of arcs in terms of Lagrangian parameters differ from the expressions at tree-level, although the UV arcs are always the same.
Now the method of Ref.~\cite{Caron-Huot:2021rmr}, which was not designed for handling loop corrections, diverges in the IR, because it involves $n~\geq~1$ dispersion relations evaluated in the forward limit. 
For illustration, we could imagine regularising these with an IR cutoff $|t|>\mu_{IR}^2$ and we would find that gravity loops enter dispersion relations with effects of order $O(\kappa^4 M^{10}/\mu^2_{IR})$ in $d=6$ or even $O(\kappa^4 M^{8}/\mu^4_{IR})$ in~$d=4$.
In $d=4$ this polynomial behaviour in $\mu_{IR}$ behaves much worse than the logarithmic ``negativity'' effects discussed in Refs.~\cite{Caron-Huot:2021rmr,Henriksson:2022oeu}.
So, in what follows we will abandon the approach of Ref.~\cite{Caron-Huot:2021rmr}.

Using FT and CS dispersion relations,  we take a perturbative approach around the tree-level bounds discussed in the previous section \ref{sec:treelevelreuslts}. For this, we rely initially on the assumption that, in the leading approximation, the functionals $f(p)$ that extremise the couplings at tree-level   are unchanged by loop effects -- we  discuss this in more detail in Appendix~\ref{sec:HOeff}.

We integrate these functionals against the 1-loop contributions, 

\begin{equation}\label{looptotree}
    \int_0^{p_{max}} \dd p\,f(p)\, \delta a_0^{CS/FT} (p)\,
\end{equation}
and then add these to the tree-level result to obtain the 1-loop corrected result --  in the FT case we have $p_{\text{max}}^2=M^2(\sqrt{17}-~1)/8$, while in the CS case we have $p_{max}=M/\sqrt{3}$.

\begin{figure}[H]
    \centering
    \includegraphics[width=0.6\linewidth]{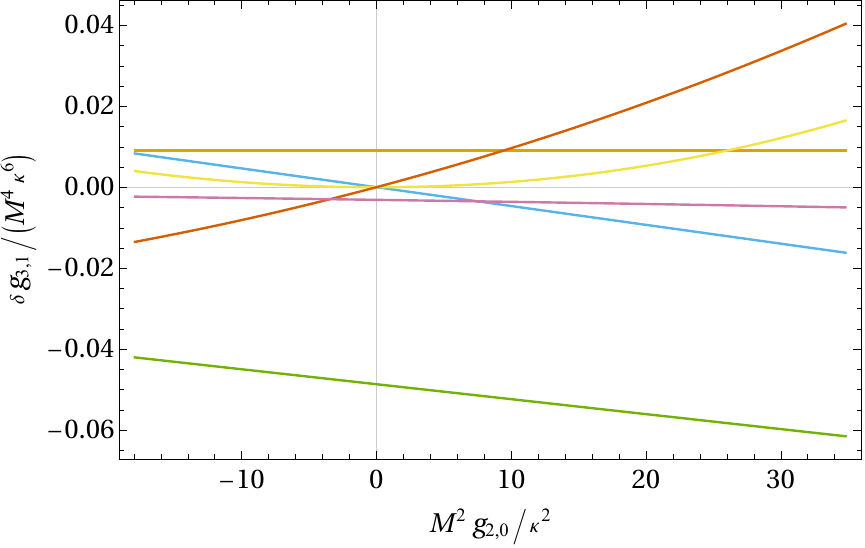}
        \hspace*{1.5cm}\includegraphics[width=0.6\linewidth]{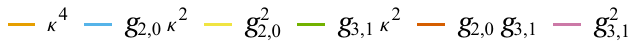}
        \includegraphics[width=0.6\linewidth]{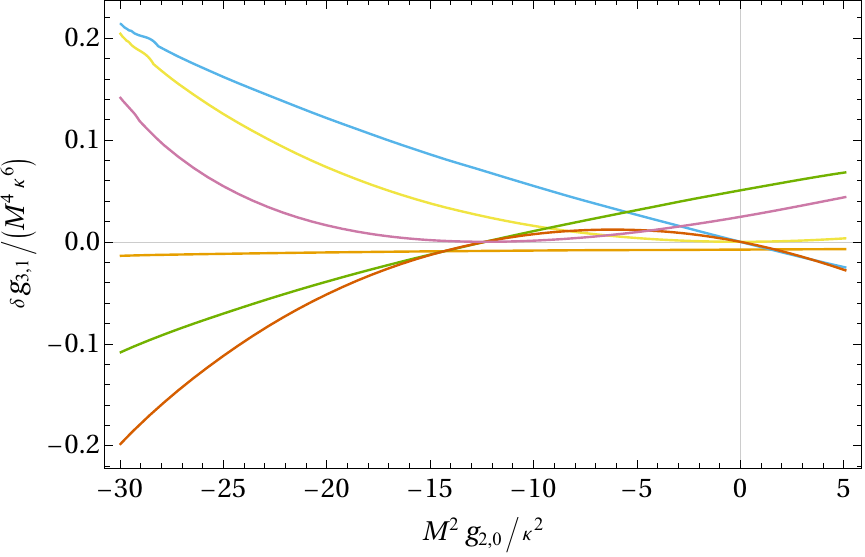}
    \caption{{\it Loop-correction to the tree-level bound, obtained with the CS method on $g_{3,1}$ varying $g_{2,0}$,
    for various loop contributions evaluated on the boundary of the allowed region, in $d=6$.
    The vertical axis has the shift in $M^4 g_{3,1}/\kappa^2$ normalised to the $\kappa^4 M^8$ scaling of a gravity loop. 
    The upper (lower) plot shows the correction to the upper (lower) bound. }}
\label{fig:comparison}
\end{figure}

The results of this analysis are shown in Fig.~\ref{fig:comparison} for CS dispersion relations, with the FT approach providing  similar results (see Appendix \ref{app:asy}). 
This figure shows the most relevant effects discussed in section~\ref{sec:IRE}, in particular effects of order $O(\kappa^4,\kappa^2 g_{2,0},g_{2,0}^2$, $\kappa^2 g_{3,1}, g_{2,0}g_{3,1},  g_{3,1}^2)$, organised here in terms  of relevance of their dimensionality.
These effects are evaluated on a point in the boundary, labelled by the value of $g_{2,0}M^2/\kappa^2$ on the horizontal axis. The normalisation of the vertical axis differs from Fig.~\ref{fig:wedge_results}, because it carries units of the gravity loop. This means that, to obtain the relative shift in $g_{3,1}/\kappa^2$ we have to first chose the size of  gravity loop effects. To guarantee perturbativity of the loop expansion these will have to be smaller than tree-level effects, $\kappa^4 M^8/(4\pi)^3 \ll \kappa^2 M^4$.

The inherently multi-scale nature of this problem makes the result non-trivial.
While the overall magnitude of these effects is governed by the scale $M$ and the size of the gravity loop, the smearing in $p$ introduces a smaller scale.
Loops involving different coefficients involve different powers of either scale and lead to effects of different size, as can be seen in Fig.~\ref{fig:wedge_results}.

The sum of all these effects is displayed in Fig.~\ref{fig:gsandman} in Appendix~\ref{app:asy}, where we see that
 FT and CS methods produce very similar results: the small differences can be traced to the different position of the tree-level boundary as discussed above. This is an important test, given   that the dispersion relations, their IR contributions, and the numerics follow completely different paths.

\subsection{A Consistent Perturbative Approach}
\label{sec:consistent_pert_approach}

An important message conveyed by Fig.~\ref{fig:comparison} is that contributions from less relevant operators, like $g_{3,0}$, appear comparable or even dominant over more relevant contributions like $\kappa^4$ in some points at the boundary. 
This is a rather generic consequence of the fact that we study loop effects around \emph{extremal} tree-level amplitudes. 
As discussed already in Refs.~\cite{Englert:2019zmt,Bellazzini:2020cot,Caron-Huot:2020cmc}, tree-level bounds tend to saturate the EFT expansion, meaning that on the bound all the coefficients, in units of the cutoff energy, have comparable size.  This is not a problem for tree-level bounds, because each coefficient can be treated almost individually as appearing only in dispersion relations with a given number of subtractions. Beyond tree-level, however, unitarity forces all couplings to enter the discontinuity, and also the arcs. So, for extremal amplitudes, it is possible that even though the loop expansion is perfectly under control, all coefficients would have to be taken into account: the EFT expansion fails and poses a problem of calculability.

Extra assumptions must be introduced  to keep the EFT expansion under control, while still working at finite couplings.
One such possibility, which preserves all physical properties of the amplitude and at the same time is in principle testable in IR experiments, is to focus on theories where the less relevant couplings are small -- see Ref.~\cite{Contino:2016jqw} for a broader discussion of this aspect inspired by  phenomenological requirements, and also Ref.~\cite{EliasMiro:2022xaa} for application of a similar condition to non-perturbative amplitudes.
For instance, assuming a small value for ${g_{2,0} M^4}/{\kappa^2} $ allows for significant gravitational effects while suppressing contributions from $g_{2,0}$. Likewise, the smallness of \( g_{3,1} \), \( g_{4,0} \), and higher-order terms shall also be assumed.  A key question is how to efficiently impose this assumption.

In the absence of gravity, assuming a small tree-level $g_{2,0}$ would be enough to ensure that all higher couplings are small. As discussed in section~\ref{sec:HOtree}, however, with gravity $g_{2,0}$ or $g_{3,1}$ can vanish without implying an inconsistency. Instead, the first condition that  we can impose which leads to  smallness of all higher order terms, is,\footnote{To study theories with seizable $g_{4,0}$ one could instead impose $g_{6,0}M^12/\kappa^2=\epsilon$; a condition on $g_{5,1}$ would not be sufficient to ensure convergence of the higher order terms.}
\begin{equation}\label{eq:leadingcond}
    \frac{g_{4,0}M^8}{\kappa^2}\leq\epsilon\,,
\end{equation}
with small enough $\epsilon$.  From the bounds of section~\ref{sec:HOtree}, we know that \eq{eq:leadingcond} implies $g_{n,q}M^{2n}{\kappa^2}\lesssim\epsilon$.\footnote{Loop effects will introduce departures from monotonicity, see Ref.~\cite{Bellazzini:2020cot}. These will be of the size of a non-divergent loop factor, hence small. In turn, such departures propagate into loops of $g_{4,0}$ as a 2-loop effect.}
This, in turn, implies that loops involving all higher  EFT coefficients are negligible.

\begin{figure}
    \centering
    \includegraphics[width=0.8\linewidth]{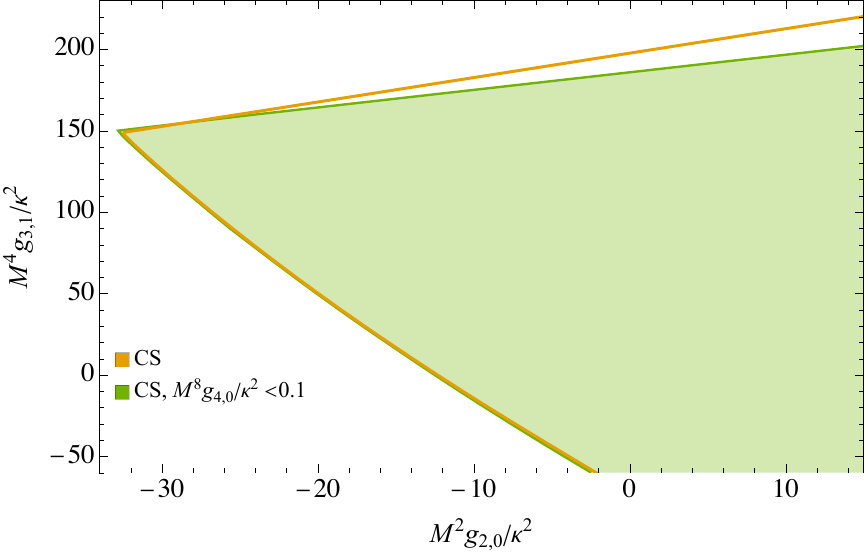}
    \caption{{\it    The tree-level bound for CS dispersion relations in $d=6$. In orange the same as  in Fig.~\ref{fig:wedge_results}. In green,  with the additional condition \eq{eq:leadingcond} with $\epsilon=0.1$, with the procedure described in Appendix \ref{app:HO}.}}
    \label{fig:g4fixed}
\end{figure}

The condition \eq{eq:leadingcond}, in fact, is not satisfied by all points in the tree-level quadrant allowed by positivity in Fig.~\ref{fig:wedge_results}. In Fig.~\ref{fig:g4fixed} we update this result to include the condition \eq{eq:leadingcond} with $\epsilon=0.1$. 
 It is immediately noticeable that the slope of the upper bound is different in this case, as expected from EFT results without gravity - we have checked that for large values of positive $g_{2,0}$ the curves asymptote to the slopes  expected in the absence of gravity.

\begin{figure}[H]
    \centering    
\includegraphics[width=0.8\linewidth]{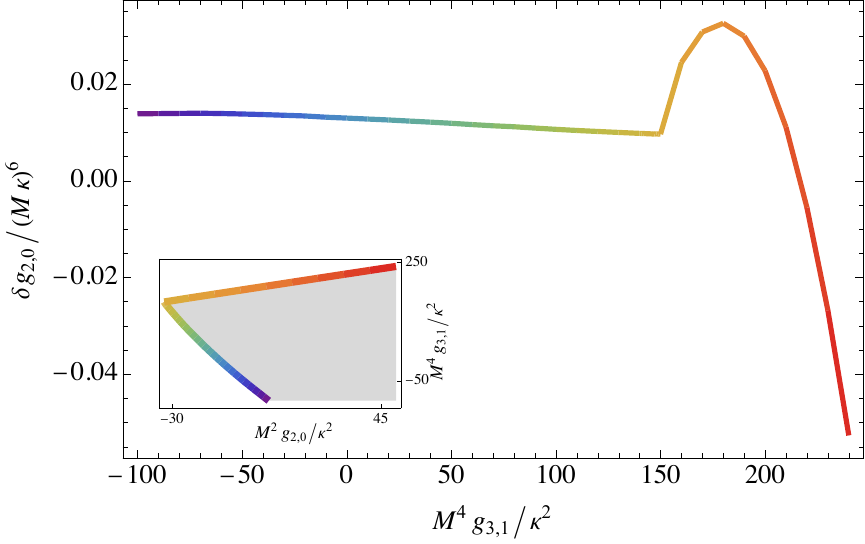}
    \caption {{\it Loop correction to the CS bounds on $g_{2,0}M^2/\kappa^2$ of Fig.~\ref{fig:g4fixed}, in units of gravity loops $\kappa^4 M^8$, at every point  $g_{3,1} M^4/\kappa^2$. The colour scheme maps each point on the tree-level bound in the inset to the associated correction. We impose $g_{4,0}M^8/\kappa^2 \leq 0.1$ in order to suppress loops from higher Wilson coefficients.}}
\label{fig:rainbow_g4fix}
\end{figure}

Now, the boundary of Fig.~\ref{fig:g4fixed} provides a robust platform on which to discuss the size of loop effects, consistent with a perturbative loop expansion \emph{and} a perturbative EFT expansion.
On the boundary of this figures, loops are entirely dominated by effects involving
only the couplings $\kappa^2, g_{2,0}$ and $g_{3,1}$, which we have computed above.
We show their sum in Fig.~\ref{fig:rainbow_g4fix}
 colour-coded in such a way to match the corresponding point on the tree-level bound, shown in the inset (notice that the axes are inverted with respect to Fig,~\ref{fig:comparison}, so that the entire boundary can be represented on the  same figure). 
The kink in the size of loop effects is of course located at the same position of the kink in the tree-level boundary. Corrections to the lower bound are much smaller than the upper bound.

\FloatBarrier %%%%TO REMOVE IF THERE IS TOO MUCH WHITE SPACE, THIS PLACES THE FIGURES BEFORE THE NEW SECTION
%%%%%%%%%%%%%%%%%%%%%%%%%%%%%%%%%%%%%%%%%%%%%%

\section{Conclusions and Outlook}

We have discussed positivity bounds from dispersion relations in the EFT of a spin-0 particle coupled to gravity. We have worked at finite IR couplings  and focussed on effects from loops of the massless particles.

Amusingly, our most important result  is  that loop effects are calculable and small -- under the right circumstances.
The path to this conclusion, however, has taught us several lessons. 
Discontinuities from a box-diagram introduce singularities in dispersion relations at small, fixed-$t$.
We have compared different methods to obtain the bounds from such dispersion relations, and found which ones are immune to these singularities: the approach of \cite{Beadle:2024hqg}
and the manifestly crossing-symmetric one \cite{Sinha:2020win,Li:2023qzs}. The methods are completely different from each other and they thus offer an important check of our results, showing agreement within our computational abilities.

With these, we have addressed the question of how much loop effects modify tree-level bounds. Alas, tree-level extremal theories tend to saturate the validity of the EFT expansion. This is an important show-stopper given that at loop level \emph{all} EFT couplings enter simultaneously dispersion relations. We have identified \eq{eq:leadingcond} as the simplest  requirement which is physical, IR testable and guarantees that only a finite number of effects must be taken into account. Then, our computation is robust, and loop-effects are indeed of the expected size, as shown in Fig.~\ref{fig:rainbow_g4fix}.
Higher-loop effects would bear no surprises, beyond the obvious quantitative refinements of the perturbative expansion.

In true Confucian fashion, the effects may be small, but the changes needed to introduce them are significant. Indeed, the combination of the IR-safe approaches \cite{Beadle:2024hqg,Sinha:2020win} and the convergence condition \eq{eq:leadingcond}, imply $O(1)$ changes in the bounds w.r.t. the orginal approach of Ref.~\cite{Caron-Huot:2021rmr}. Moreover, these changes are evident already at tree-level, as we have shown in~Fig.~\ref{fig:g4fixed}.

Finally, we identified the first running coefficients: \( g_{2,0} \) for \( d \leq 4 \), \( g_{3,1} \) for \( 4 < d \leq 6 \), and \( g_{4,0} \) for \( 6 < d \leq 8 \). This is significant because the running of a relevant operator dominates over less relevant ones at low energy. In dispersion relation bounds, this implies that for \( d = 5,6 \), the (running) coefficient \( g_{4,0}(M) \), along with all less relevant terms, receives polynomially growing corrections at small \( M \). For \( d \geq 7 \), this behavior begins at \( g_{6,0}(M) \). 
The bounds on these coefficients are determined by forward dynamics, and  indicate that Wilson coefficients can become negative, as discussed in Ref.~\cite{Bellazzini:2020cot} -- the sign-definiteness being  a direct consequence of unitarity in the EFT.

There are many further questions worth pursuing. 
First of all it would be interesting to refine the approach of Ref.~\cite{Beadle:2024hqg} with an all-order formula. This might exhibit better numerical stability and would facilitate the comparison with the crossing symmetric approach. 
It would also be useful as a complementary tool to crossing symmetric dispersion relations, based on different assumptions about the amplitudes analytic structure.

Another important question would be the direct exploration of healthy UV completions involving gravity. In particular, understanding if the upper bound $|t|< M^2/2$ implied by the loop-resilient approaches bears any deeper meaning in terms of assumptions on the underlying theories that satisfy these dispersion relations. Are there any theories that satisfy our bounds and not those of Ref.~\cite{Caron-Huot:2021rmr}?
Conversely, can we exploit these different $t$-ranges to exclude unwanted theories from the UV spectrum? Smearing at values $|t|>2M^2$ could rule out certain UV theories with accumulation points/double poles, like the $stu$ amplitude of \eq{eq:stu}.

Some of the aspects discussed here are expected to arise also in the theory of electromagnetism coupled to pions in singly-subtracted dispersion relations~\cite{Albert:2022oes,Fernandez:2022kzi,Albert:2023seb}. In particular, it would be interesting to see how constraints involving anomalies~\cite{Albert:2023jtd,Ma:2023vgc,Dong:2024omo} are impacted by the finite-$N_c$ effects we discuss, given that the anomaly itself is loop generated. 

More generally, it would be interesting to establish a  solid bridge between the positivity program on  weakly coupled gravitational theories, and  fully non-perturbative approaches to the S-matrix bootstrap, as in \cite{Guerrieri:2021ivu,EliasMiro:2022xaa}. We believe our work consitutes an important step in this direction.

\section*{Acknowledgments}
We thank B. Bellazzini, J.~Elias-Miro, M.~Gumus, Y.~Huang, J.~Parra-Martinez, L.~Rastelli, F.~Sciotti, A.~Sinha and A. Zhiboedov for important discussions. 
A particular thank goes to D.~Kosmopoulos for his help with graviton amplitudes
and to C.~Chang and J.~Parra-Martinez for important  exchanges  related to their  work in progress. The work of
C.B. was supported by Swiss National Science Foundation (SNSF) Ambizione grant PZ00P2-
193322. G.I. is supported by the US Department of Energy under award number DE-SC0024224, the Sloan Foundation and the Mani L. Bhaumik Institute for Theoretical Physics.
G.I., D.P., S.R. and F.R. have been supported by the SNSF under grants no. 200021-205016 and PP00P2-206149.
The work of F.S. was supported by the Simons Investigator Award No. 827042 (P.I.: Surjeet Rajendran) and the Simons Investigator Award No. 144924 (P.I.: David E. Kaplan).

%%%%%%%%%%%%%%%%%%%%%%%%%%%%%%%%%%
%\newpage

\appendix

\section{Arcs}
\label{app:arcs}
In this Appendix we list the expressions for the arcs' corrections from loops of order $O(\kappa^4)$, for the CS case of Sec. \ref{sec:loopscsdr}.

\begin{align}
\label{eq:csGRpiece}
        \delta a_0^{\text{CS}} &= -\frac{173 \bar{p}^2 \log (\mu )}{2240 \pi ^3}-\frac{3 \bar{p}^2 \text{Li}_2\left(-\frac{1}{\bar{p}^2}\right)}{64 \pi ^3}+\frac{\log \left(\frac{1}{\bar{p}^2}+1\right)}{64 \pi ^3 \bar{p}^2} \nonumber \\
        &+\frac{1}{1881600 \pi ^3}\left(\bar{p}^2 \left(-\frac{9800 \bar{p}^2 (\bar{p}^2 (\bar{p}^2+3)+3) \log (8) \log (32)}{(\bar{p}^2+1)^3}+34300 \pi ^2-76529\right)-148820\right)\nonumber\\
        &-\frac{3 \sqrt{\frac{4}{\bar{p}^2+1}-3} \tanh^{-1}\left(\sqrt{\frac{4}{\bar{p}^2+1}-3}\right)}{32 \pi ^3 (\bar{p}^2+1)^3}-\frac{1409 \bar{p}^2 \coth ^{-1}(2 \bar{p}^2+1)}{13440 \pi ^3}\nonumber\\
        &+\frac{1}{215040 \pi ^3 \bar{p}^2 (\bar{p}^2+1)^3}\left[875 \sqrt{\frac{4}{\bar{p}^2+1}-3}\, \bar{p}^6 \log (4 \bar{p}^2 (\bar{p}^2+1))\right. \nonumber\\
        &+2485 \sqrt{\frac{4}{\bar{p}^2+1}-3}\  \bar{p}^6 \log \left(\bar{p}^2-\sqrt{(1-3 \bar{p}^2) (\bar{p}^2+1)}+1\right)\nonumber\\
&-4235\sqrt{\frac{4}{\bar{p}^2+1}-3} \ \bar{p}^6 \log \left(\bar{p}^2+\sqrt{(1-3 \bar{p}^2) (\bar{p}^2+1)}+1\right)\nonumber\\
&+3360 \bar{p}^4 \left[(\bar{p}^2+1)^3 \left[\log^2\left(\bar{p}^2-\sqrt{(1-3 \bar{p}^2)(\bar{p}^2+1)}+1\right)+\log^2\left(\bar{p}^2+\sqrt{(1-3 \bar{p}^2) (\bar{p}^2+1)}+1\right) \right. \right. \nonumber\\
&\left. -7 \log \left(\bar{p}^2+\sqrt{(1-3 \bar{p}^2) (\bar{p}^2+1)}+1\right) \log\left(\bar{p}^2-\sqrt{(1-3 \bar{p}^2) (\bar{p}^2+1)}+1\right)+5 \log (4 \bar{p}^2) \log(\bar{p}^2+1)\right] \nonumber\\
&+\left. \log (32) \log \left(\frac{2 \bar{p}^2}{\bar{p}^2+1}\right)+5  (\bar{p}^4 (\bar{p}^2+3)+3\bar{p}^2) \log (2) \log \left(\frac{4 \bar{p}^2}{\bar{p}^2+1}\right)\right] \nonumber \\
&-1001 \sqrt{\frac{4}{\bar{p}^2+1}-3} \bar{p}^4 \log (4 \bar{p}^2 (\bar{p}^2+1))+11081 \sqrt{\frac{4}{\bar{p}^2+1}-3} \bar{p}^4 \log \left(\bar{p}^2-\sqrt{(1-3 \bar{p}^2) (\bar{p}^2+1)}+1\right)\nonumber \\
&-9079\sqrt{\frac{4}{\bar{p}^2+1}-3} \bar{p}^4 \log \left(\bar{p}^2+\sqrt{(1-3 \bar{p}^2)(\bar{p}^2+1)}+1\right)-3360 \sqrt{\frac{4}{\bar{p}^2+1}-3} \log (2 \bar{p}^2)\nonumber\\
&\left.+3360 \sqrt{\frac{4}{\bar{p}^2+1}-3} \log \left(-\bar{p}^2-\sqrt{(1-3 \bar{p}^2) \
(\bar{p}^2+1)}+1\right)\right].
\end{align}

\section{Bounds from Smearing}\label{app:smearing}
Here we review the FT algorithm of~\cite{Beadle:2024hqg} (in turn based on \cite{Caron-Huot:2021rmr}) and adapt it to CS dispersion relations. We also explain details of how the semi-definite optimisation program is built and how we treat the logarithms appearing in IR arcs. 
In this Appendix, for simplicity, we work in units of the cutoff, $M^2=1$.

We write a generic smearing function $f(p)$ as,
\begin{equation}\label{eq:measure} 
    f(p)= p^{\alpha}\,\sum_{j=0}^{j_{\text{max}}} c_j p^{\,j},
\end{equation}
where a constant overall factor $p^\alpha$ is added to integrate to a finite value on the gravity pole -- $\alpha$ will be fixed later in the procedure. 
We define the vector $\vb*{W}$ in the UV as
\begin{equation} W_{j,\;\ell}(s) =
     \int_0^{1/\sqrt{3}} \dd p \,p^{j+\alpha}\;  I_{\ell,0}(s,p).
\end{equation} 
While in the IR, we define the vector $\vb*{V}$ as
\begin{equation} 
    V_j =  \int_0^{1/\sqrt{3}} \dd p\,  p^{j+\alpha}\,a_{0}^{CS}(p).
\end{equation} 
Then, positivity bounds can be written as 
\begin{equation}\label{eq44app}
    \sum_{j=0}^{j_{\rm{max}}} c_j W_{j,\;\ell}(s) \geq 0, \quad \forall \,s,
    \; \ell, \; \implies  \sum_{j=0}^{j_{\rm{max}}} c_j V_j \geq 0.
\end{equation} 
The coefficients $c_j$ can be varied, in search of an optimal function which minimises or maximises the bound.
We therefore have a target vector $V_j$ and a vector to optimise $c_j$ tin order to obtain $\sum c_j V_j
\geq 0$, subject to some constraints $W_{j,\;\ell}(s) \succcurlyeq0$.  This defines a semi-definite optimisation problem.

In principle positivity should be imposed for all values of $\ell$. 
In practice, we work with a finite $\ell\leq\ell_{\text{max}}$ and increasing its value until the bound stabilises.

The finite range $\ell\leq \ell_{\text{max}}$ means less UV conditions and hence artificially stronger bounds.  As described below \eq{smeardingexp} this can be complemented with $\ell\to \infty$  information, through the \emph{finite impact parameter} limit to gather information at large values of $\ell$.
This corresponds to $s\to \infty$, $\ell \to \infty$ with   $b=2\ell/\sqrt{s}$ fixed. In this limit the Gegenbauer polynomials give,
\begin{equation}
    \lim_{m\to \infty}\mathcal{P}_{mb/2}\left(\sqrt{\frac{m^2-3p^2}{m^2+p^2}}\right) = \frac{\Gamma\left(\frac{d-2}{2}\right)}{\left(\frac{bp}{2}\right)^{\frac{d-4}{2}}} J_{\frac{d-4}{2}}(bp)\,,
\end{equation}
with $m\equiv \sqrt{s}$. 
In this limit the dispersion relations can be integrated exactly into fixed-impact-parameter relations,
\begin{equation}\label{eq:fixed-b}
    W_j(b)= \Gamma\left(\frac{d-2}{2}\right)\,\int_0^{1/\sqrt{3}} \dd p \,p^{j+\alpha}\frac{J_{\frac{d-4}{2}}(bp)}{\left(bp/2\right)^{\frac{d-4}{2}}}= \left(\frac{1}{3}\right)^{(j+\alpha+1)/2} \frac{{}_1F_2\left(\frac{j+\alpha+1}{2}; \frac{d-2}{2},\frac{j+\alpha+3}{2};-\frac{b^2}{12}\right)}{j+\alpha+1}.
\end{equation}
This leads to more constraints on the function $f$, complementary to fixed-$\ell$. In particular we  impose \eq{eq:fixed-b} at fixed values of $b$ and, following  Ref. \cite{Caron-Huot:2021rmr}, we  demand that is positive for large values of $b$, where,
\begin{equation}
    \frac{{}_1F_2\left(\frac{j+\alpha+1}{2}; \frac{d-2}{2},\frac{j+\alpha+3}{2};-\frac{b^2}{12}\right)}{j+\alpha+1} \sim A(b) + B(b)\cos(\phi) +C(b)\sin(\phi),
\end{equation}
with $\phi$ defined as the argument of the oscillatory terms, which depends on $b$, and $A,B,C$ obtain at a  certain order in the $1/b$ expansion. 

Positivity for all $\phi$ requires,
\begin{equation}
    \begin{pmatrix}
    A(b) + B(b) & C(b)\\
    C(b)&A(b) - B(b)
    \end{pmatrix}\succcurlyeq0.
\end{equation}

For these expressions to be  polynomial in $b$ and to satisfy the condition $4-d+\alpha\leq 0$ implied by Bochner's theorem (see below \eq{smeardingexp}), we demand $\alpha = \frac{d-3}{2}$. Then, up to an overall factor, we are left with a matrix of polynomials in $b$, which can be treated with the usual techniques of semi-definite optimisation.

\begin{figure}[H]
    \centering
    \includegraphics[width = .7\linewidth]{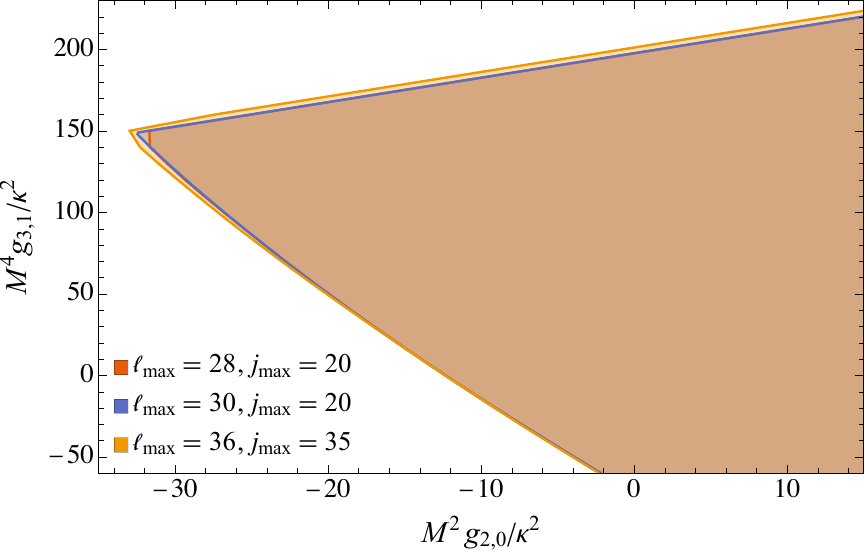}
    \caption{\it Bounds in the CS case, for various values of $\ell_{\rm max}$, the number of spins, and $j_{\rm max}$, the number of elements in the basis. The bound becomes stable with $\ell_{\rm max}\sim 30$ and $j_{\rm max}\sim 30$.}
    \label{fig:Jcheck}
\end{figure}

For the bounds described in the main text, we have used,
\begin{equation} \label{appCtarget}
    V_j =   \int_0^{1/\sqrt{3}} \dd p\,p^{\alpha + j}  \left( \frac{\kappa^2}{p^2}+ 
    g_{2,0} + g_{3,1} p^2 + \mathcal{O}(\log(p^2))\,\right)\,, 
\end{equation} 
and optimised the value of $g_{3,1}/\kappa^2$ for a fixed value of $g_{2,0}/\kappa^2$ using the software {\tt sdpb}~\cite{Simmons-Duffin:2015qma,Landry:2019qug}.
We utilise multiple values of $\ell\leq\ell_{\text{max}}$ and  discretise in $\sqrt{s} \in [1,\infty]$ in the CS case, by defining 
    $\sqrt{s}=\frac{1}{1-x}$
with $x\in [0,1-\delta x]$ sampled in steps $\delta x=0.1$. We have checked that smaller values of $\delta x$ do not change quantitatively the results.
Furthermore we show how the plot changes when increasing $\ell_{\rm max}$ and $j_{\rm max}$ in Fig. \ref{fig:Jcheck}. Changing these parameters does not modifies the plot qualitatively.

We built a matrix with rows made out of $1\times1$ matrices for each value of $\ell$ and $x$, and a column for each element of the basis of polynomials.
For the optimisation procedure a normalisation choice for $f(p)$ is required.
\begin{equation}
 \int_0^{1/
\sqrt{3}}\dd p \; f(p) \,p^2 =  \sum_{j=0}^{j_{\rm{max}}}c_j \int_0^{1/
\sqrt{3}}\dd p \;  p^{\,j+\frac{d+1}{2}} = \pm1, 
\end{equation} 
where the $-$ sign gives the upper bound and the $+$ the lower bound.

\subsection{Loop order functionals}\label{sec:HOeff}

\begin{figure}[H]
    \centering
    \includegraphics[width=0.75\linewidth]{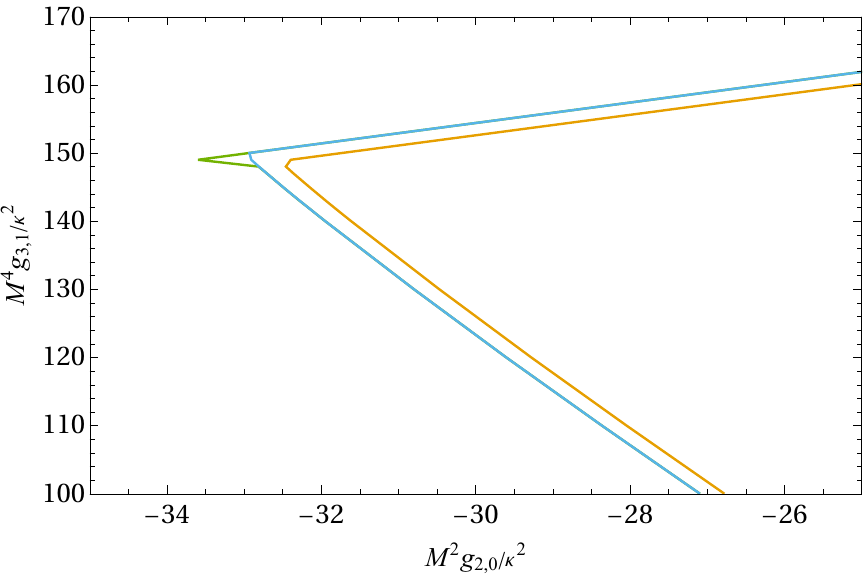}
    \caption {{\it Comparison, in the CS case, between the perturbative expansion used in Sec.\ref{sec:BoundsMod} (blue), and a direct approach of including the loops in the semidefinite optimisation problem (green) both for the gravitational loop only, with fixed $\kappa^2/(4\pi)^3 = 0.1M^{-4}$. In orange the tree-level bound.}}
    \label{fig:loopExplicit}
\end{figure}

In our perturbative approach we have employed the tree-level extremal functionals $f(p)$ to compute the loop-level contributions via \eq{looptotree}. In principle, dispersion relation with loop effects might be extremised by other ``loop-level''  functionals.
Fig.~\ref{fig:loopExplicit}  shows the difference between using tree-level and loop-level functionals, in the context of CS dispersion relations. For clarity we have limited the analysis to  loops involving only gravity, with $\kappa^2/(4\pi)^3 = 0.1M^{-4}$.

  The deviations are most notable near the kink, with the rest of the bound being unaffected.
This difference can be explained by the fact that the $g_{3,1}$ position of the kink is not
captured by the tree-level functionals.

\section{Bounds with fixed higher order Wilson coefficients}\label{app:HO}

As described in section~\ref{sec:consistent_pert_approach}, to ensure  control over the EFT expansion into loop effects, we demand the condition \eq{eq:leadingcond}. This is reflected in Figs.~\ref{fig:g4fixed} and~\ref{fig:rainbow_g4fix}. For practical reasons, it is simpler to impose -- beside $M^6 g_{4,0}/\kappa^2=\epsilon$ -- also $M^8 g_{5,1}=M^{10} g_{6,2}=\epsilon\kappa^2$, although smallness of $g_{5,1},g_{6,2}$ is normally implied by the condition on $g_{4,0}$.

The reason for this is that we impose these conditions
by adding to the 0-th CS arc $a_0^{\rm CS}$, also $p^4 a_1^{\rm CS}$,
\begin{equation}
    V_j = \int_0^{1/\sqrt{3}}\dd p\,p^{j+\alpha}\left[a_0^{\rm CS}(p)+p^4a_1^{\rm CS}(p)\right],
\end{equation}
where the first arc in the IR reads,
\begin{equation}
    a_1^{\rm CS}(p) = g_{4,0} + g_{5,1}p^2 + g_{6,2}p^4.
\end{equation}
By entering specific values of each parameter in the objective function we fix their values. In this way we obtain Fig.~\ref{fig:g4fixed}). 
The lower bound and the kink are unchanged, while the slope of the upper bound changes from $3/2$ to $\sim1.08$. This is equally expected, since the scalar UV completion is excluded by our choice of Wilson coefficients. We use all the same parameters as for Fig.~\ref{fig:wedge_results}, with $\ell_{\rm max} = 30$, and 20 elements in the basis of polynomials.

\section{Fixed-$t$ versus Crossing Symmetric Dispersion Relations}\label{app:asy}
\begin{figure}
    \centering
\includegraphics[width=0.75\linewidth]{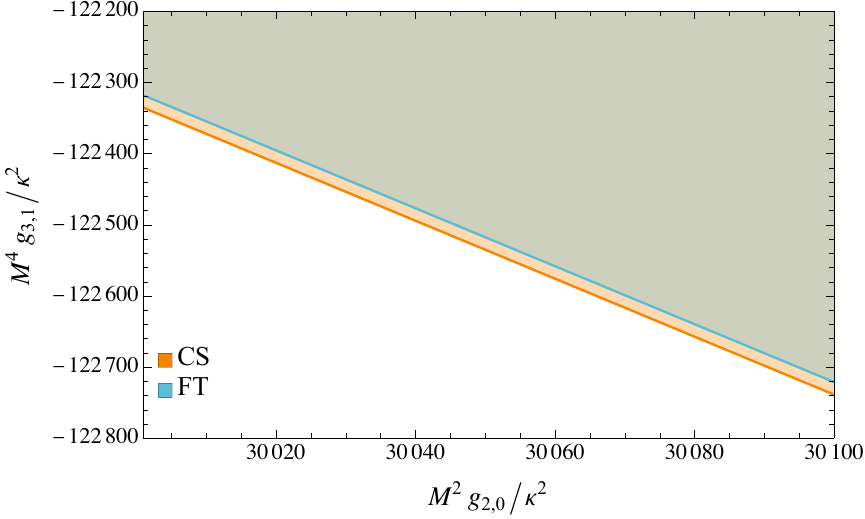}
   \caption{{\it Asymptotic behaviour of the lower bound in the regime where gravity is negligible, i.e. $\kappa^2\ll~M^{2(n-1)}\left|g_{n,q}\right|$. The slope here gives $M^2 g_{3,1}/g_{2,0}\geq -4.07$, which is consistent with the tree-level bound without gravity. Both CS and FT methods coincide with this slope value.}}
\label{fig:asymptotics}
\end{figure}

As referenced in section \ref{sec:treelevelreuslts}, in the region where the effects of gravity are small, $\kappa^2\ll~M^{2(n-1)}\left|g_{n,q}\right|$, the asymptotic of the lower boundary reproduces a slope of $\sim -4.07$, which is indeed the lower bound on the ratio $M^2 g_{3,1}/g_{2,0}$ in the absence of gravity. We show this in Fig.\ref{fig:asymptotics}. The upper  slope reaches the asymptotic value of $3/2$ already close to the tip, therefore is not shown here.

\begin{figure}
    \centering
    \includegraphics[width=0.75\linewidth]{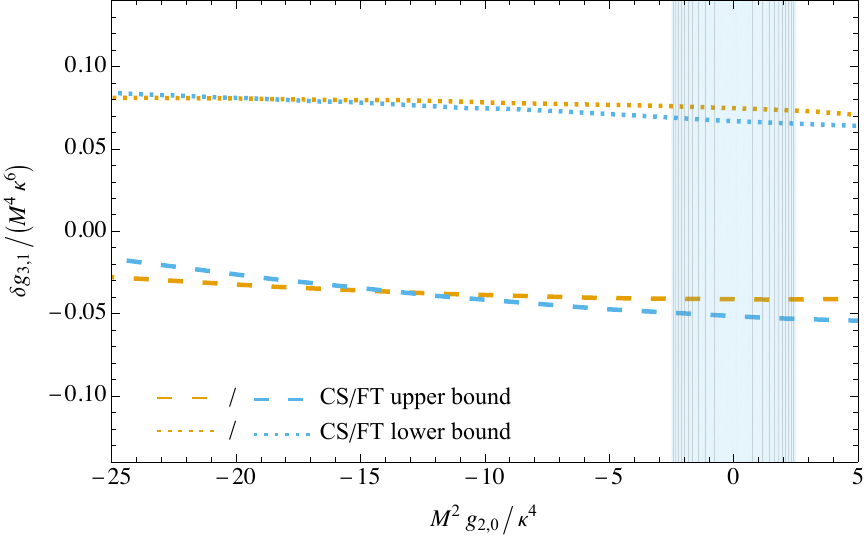}
    \caption{\it Correction to $\delta g_{3,1}$ for each point of $g_{2,0}$. The CS (FT) method is plotted in orange (blue), while the upper (lower) bound is dashed (dotted). The upper bound gets corrected less than the lower bound.}
    \label{fig:gsandman}
\end{figure}

A further comparison between CS and FT methods is given by the correction to the bounds on $g_{2,0}$ and $g_{3,1}$ in the presence of gravity and all loops, which we show in Fig.~\ref{fig:gsandman}. 
These corrections are displayed as a deviation $\delta g_{3,1}/(M^4 \kappa^6)$ in terms of $g_{2,0}M^2/\kappa^4$ -- the same as Fig.~\ref{fig:comparison} but opposite than Fig.~\ref{fig:rainbow_g4fix}.

\newpage
\bibliographystyle{JHEP}
\bibliography{draft}

% Fleas: Adam had'em
\end{document}